\documentclass[11pt]{article}
\usepackage{amsmath}
\usepackage{graphicx}
\usepackage{epsfig}
\usepackage{float}
\usepackage{makeidx}
\usepackage{bbm}
\usepackage{hyperref}
\usepackage{verbatim}
\usepackage{caption}
\usepackage{verbatim}
\usepackage[utf8]{inputenc}
\usepackage[left=2.5cm,right=2.5cm,top=1.5cm,bottom=3cm]{geometry}
\usepackage{enumerate}
\usepackage{amssymb,amsmath,amsthm,amsfonts}
\usepackage{pstricks,pstricks-add,pst-math,pst-xkey}
\usepackage{makeidx}
\usepackage{cancel,tikz}\usepackage{float}
\usepackage[missing={Father rev: f17b7ec 2020-07-08}]{gitinfo}
\usepackage{graphicx}
\usepackage[font=small,labelfont=bf]{caption}
\usepackage{subfigure}
\usepackage{tikz}
\usetikzlibrary{calc,patterns}
\usetikzlibrary{math}
\usetikzlibrary{intersections}
\usetikzlibrary{scopes}
\usetikzlibrary{shapes}
\usetikzlibrary{arrows,automata}
\newlength{\altezzautile}
\newlength{\larghezzautile}
\usepackage{pgfplots}

\newtheorem{theorem}{Theorem}

\newtheorem{lemma}{Lemma}
\newtheorem{corollary}{Corollary}

\makeindex

 \let\b=\beta

 \let\g=\gamma      
  \let\om=\omega     
 \let\s=\sigma \let\t=\tau 
\let\y=\upsilon \let\x=\xi 
  \let\G=\Gamma \let\L=\Lambda 
\let\O=\Omega   \let\Si=\Sigma

\def\\{\noindent}
\def\Z{\mathbb{Z}}
\def\N{\mathbb{N}}
\def\Zd{\mathbb{Z}^d}

\def\Z2{\mathbb{Z}^2}

\def\PP{\mathcal{P}}

\def\0{\emptyset}

\def\CC{\mathcal{C}}

\def\1{\mathbbm{1}}
\def\<{\langle}
\def\>{\rangle}
\def\be{\begin{equation}}
\def\ee{\end{equation}}
\def\<{\langle} \def\>{\rangle}
\def\PP{{\mathbb P}}
\def\0{\emptyset}
\def\be{\begin{equation}}\def\ee{\end{equation}}
\def\1{\rlap{\mbox{\small\rm 1}}\kern.15em 1}

\def\XX{{\rm X}}
\def\YY{{\rm Y}}
\def\\{\noindent}
\def\Z{{\mathbb{Z}}}
\def\ev{{\mathfrak{e}}}
\def\conplus{^{_{\;+}}_{^\leftrightsquigarrow}}
\def\conminus{^{_{\;-}}_{^\leftrightsquigarrow}}

\begin{document}

\title {The  Blume-Emery-Griffiths  model at the FAD and AD interfaces}

\author{Paulo C. Lima$^{\dag}$, Riccardo Mariani$^{*}$,  Aldo Procacci$^\dag$ and Benedetto Scoppola$^*$ \\\\
\footnotesize{$^\dag$ Dep. Matem\`atica-ICEx, UFMG, CP 702 Belo Horizonte - MG, 30161-970 Brazil}\\
\footnotesize{$^*$Dipartimento di Matematica - Universit\`a Tor Vergata di Roma, 00133 Roma, Italy}\\
\tiny{emails: {pcupertinolima@gmail.com};~{riccardo.mariani211@gmail.com};}
\tiny{aldo@ufmg.br};~{scoppola@mat.uniroma2.it}}
\date{}
\maketitle
\begin{abstract}
We analyse the Blume-Emery-Griffiths (BEG) model on the lattice $\Zd$ at the ferromagnetic-antiquadrupolar-disordered (FAD) and antiquadrupolar-disordered (AD) interfaces of parameters. In our analysis of the FAD interface we introduce a Gibbs sampler of the ground states at zero temperature, and we exploit it in two different ways: first, we perform via perfect sampling an empirical evaluation of the spontaneous magnetization at zero temperature, finding a non-zero value in $d=3$ and a vanishing value in $d=2$. Second, using a careful coupling with the Bernoulli  site percolation model in $d=2$, we prove rigorously that imposing $+$ boundary conditions,  the magnetization in the center of a square box tends to zero in the thermodynamical limit and the two-point correlations decay exponentially. Also, using again a coupling argument, we show  that the infinite volume Gibbs measure of the zero-temperature BEG exists and it is unique. In our analysis of the AD interface we restrict ourselves to $d=2$ and, by comparing the BEG model with  a Bernoulli site percolation in a matching graph of $\mathbb{Z}^2$,  we  get  a condition for the vanishing of the infinite volume limit magnetization improving, for low temperatures, earlier results obtained via expansion techniques.

\end{abstract}

\section{Introduction}\mbox{}\\

The Blume-Emery-Griffiths (BEG) model was introduced in 1971 in order to explain the superfluidity and phase transition of $He^3-He^4$ mixtures \cite{beg71} and has been extended and generalized to many other applications, among them, ternary fluids \cite{mb74,fdg77}, phase transitions in $UO_2$ \cite{g67} and $DyVO_4$ \cite{sb72}, phase changes in microemulsion \cite{ss86}, solid-liquid-gas system \cite{ls75} and semiconductor alloys \cite{dn83}.  The formal Hamiltonian of the BEG model with zero magnetic field is given by
\begin{eqnarray}\label{defbeg}
 H (\sigma)=-\sum_{x\sim y} (\sigma_x\sigma_y+\YY\sigma_x^2\sigma_y^2+\XX(\sigma_x^2+\sigma_y^2)),
\end{eqnarray}
where by the notation  $x\sim y$ we mean that $\{x,y\}$ is an unordered pair of nearest neighbors in $\mathbb{Z}^d$, $\sigma_x\in\{-1,0,1\}$ and $\XX,\YY\in\mathbb{R}$.

To understand the low-temperature properties of the model the starting point is to establish
its ground state configurations and this is done, for instance, in \cite{bl00}, where the $\XX\YY$-plane is decomposed into three regions (according to the lowest spin pair energies), namely
\begin{eqnarray*}
F &=& \{(\XX, \YY) : 1 + 2\XX + \YY > 0 \mbox{ and }   1 + \XX + \YY > 0\}\\
D &=& \{(\XX, \YY) : 1 + 2\XX + \YY < 0 \mbox{ and }  \XX < 0\}\\
A &=& \{(\XX, \YY) : 1 + \XX + \YY < 0 \mbox{ and }  \XX > 0\},
\end{eqnarray*}
called in the physics literature as ferromagnetic, disordered and antiquadrupolar,
respectively. In these regions the spin pairs with lowest energies are $\{++,--\}$, $\{00\}$
and $\{0+,0-\}$, respectively. In particular, for $(\XX, \YY)\in D$ the constant configuration
$\sigma_x=0$, for all $x$, is the only ground state. For $(\XX, \YY)\in F$ there are two ground
states, namely the constant configurations $\sigma_x=+1$, for all $x$, and $\sigma_x=-1$, for all $x$, respectively.
For $(\XX, \YY) \in A$ the model has infinitely many ground states separated into two disjoint classes. The first class is formed by those configurations $\s$  such that $\sigma_x=0$ for all  $x$ is the even sublattice of $\mathbb{Z}^d$, and the second class is the set of those configurations $\t$  such that $\tau_y=0$ for all  $y$ is the odd sublattice of $\mathbb{Z}^d$.

\begin{figure}[H]
\centering
\includegraphics[width=0.7\textwidth]{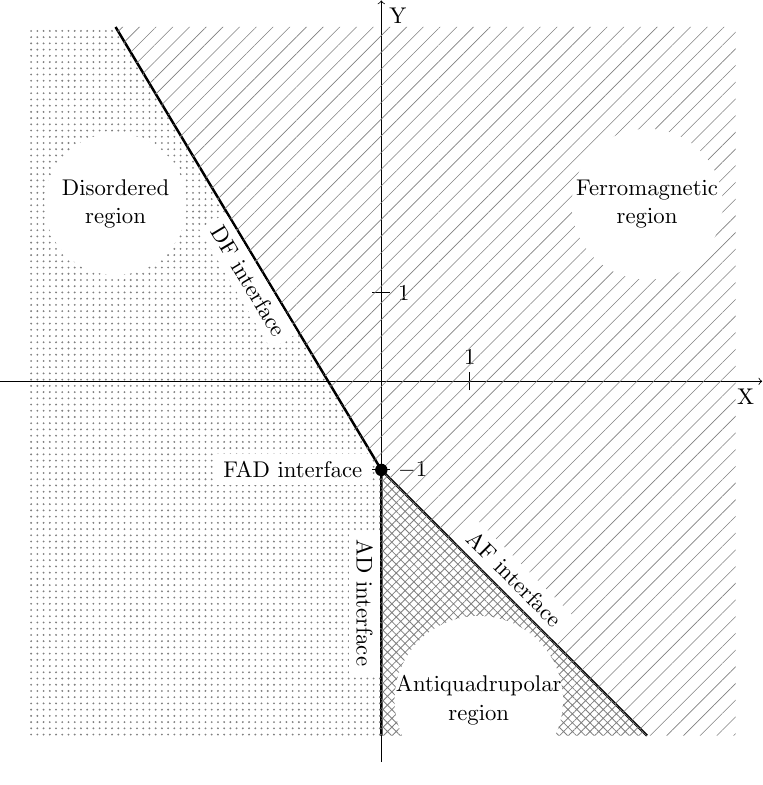
}
\caption{\label{fig:phase_diagram}  The regions $A$, $D$, $F$ and their interfaces $AD$, $DF$, $AF$ and $FAD$.}
\end{figure}

The boundaries of the above regions are the lines
$DF=\{(\XX,\YY)\,:\, 1+2\XX+\YY=0 \mbox{ and } \XX<0\}$,
$AF =\{(\XX,\YY)\,: \, \XX+\YY+1=0 \mbox{ and }  \XX>0\}$ and
$AD =\{(0,\YY)\,:\, \YY<-1\}$, and the point $\{(0,-1)\}$. In the physics literature this point is called ferromagnetic-antiquadrupolar-disordered interface and will be denoted by FAD. For the values of parameters $(\XX,\YY)$
at the $AF$, $AD$ and $FAD$ interfaces the BEG model has  nonzero residual entropies. The references  \cite{bl00,bl08,ln11} provide lower and upper bounds for the their values in two dimensions, by means of the transfer matrix method.

In the high temperature regime (i.e $\b $ small) the behavior of the model for any values of the parameters $\XX$ and $\YY$  can be
described via standard polymer-expansion techniques.
In \cite{pcl21}, by using the  Dobrushin uniqueness criterion,  it is shown the existence of subset of  $D$ for which there is  a unique Gibbs state for all temperatures, while in \cite{llp}, via cluster expansion techniques, it is exhibited another  subset of $D$ where the pressure of the system is analytic at all temperatures.
In \cite{pcl20} some correlation functions of BEG model are analysed for $(\XX, \YY)$ in the region $D$ and at the $AD$ interface and,  among other results, it is shown that for $|\YY|$ sufficiently large, depending on $d$,  the  magnetization is zero for all temperature.

For those parameters $(\XX,\YY)$  for which there are  only  a finite number of ground states, namely, in the regions $F$ and $D$ and at the interface $DF$, the low temperature description of the model is given by the usual Pirogov-Sinai theory \cite{si82,sv17}.    When $(\XX,\YY)$ are in the region $A$, where the model has an infinite number of ground states separated into two disjoint classes,  an extension of the Pirogov-Sinai theory given in  \cite{gs88} allows to prove that at low temperature there are only two translation-invariant Gibbs states (see \cite{bl00}).   The situation at the  $AF$, $AD$ and $FAD$ interfaces is quite different: besides the fact the degeneracies of their ground states are higher than the ones of the region $A$, the ground states in these interfaces do not split into a finite number of equivalence classes and the usual  Pirogov-Sinai theory and  its known extensions fail to work.  Consequently, at the  $AF$, $AD$ and $FAD$ interfaces the behavior of the model at low temperature, even at zero-temperature, is less clear.

It is worth mention that for any temperature, by spin flipping of the spins in one of the sublattices of $\mathbb{Z}^d$, the   BEG model with parameters $\XX=2$ and $\YY=-3$ (a point of the AF interface) is mapped into the three state antiferromagnetic Potts model.  Moreover, by this operation, the zero-temperature BEG model at the whole AF interface is mapped into the three-state antiferromagnetic Potts model, namely the proper three-colorings problem.
Generally,  concerning the $q$-state   antiferromagnetic Potts model (with $q\ge 3$) on a given lattice ${\cal L}$ , it is expected that there  is  a  value $q_c({\cal L})$ such that if $q<q_c({\cal L})$ the model orders at low temperature, if
 $q=q_c({\cal L})$ the model has a critical point at zero temperature, and if $q>q_c({\cal L})$ it is disordered at all temperatures (see e.g. \cite{PSG} for a lower bound on $q_c({\cal L})$ when ${\cal L}$ is a quasi-transitive infinite graph).
 For  the case ${\cal L}=\mathbb{Z}^d$ and $q=3$ it is expected that  the ordered phase is such that
 one of the two sublattice (i.e. either the  even sublattice or the odd sublattice)  is mostly occupied
 by a single state, while the spin values on the other sublattice are split equally between the remaining two states
 (the so called  broken-sublattice-symmetry (BSS) phases).
 For  the case ${\cal L}=\mathbb{Z}^2$ and $q=3$
strong theoretical arguments, which however, fall off a rigorous proof, predicts that this  model has a critical point at zero temperature \cite{ss98} ordering according the BSS phase. Rigorous proofs of the existence of the BSS phase are available only for $d$ sufficiently large, see \cite{PS} and references therein.
Finally, we also mention that the zero-temperature BEG model at the whole AD is equivalent to  the self-repulsive hard-core gas (see  \cite{ss05} for a review on  this model) with activity $2$  which, in the two-dimensional case, has been showed to be in the uniqueness phase \cite{RSTVY}.

The present paper consists of two parts. The first part (Section \ref{sec:definition}) is devoted to the analysis of the zero-temperature BEG model on the lattice $\Zd$ at  the FAD interface while in  the second part (Section \ref{7}) we consider  the low temperature  BEG model on the lattice
$\mathbb{Z}^2$ at the
AD interface.

In our analysis of the FAD interface we introduce in Sec. \ref{3} a Gibbs sampler of the ground states at zero temperature, and we exploit it in two different ways. First, we perform via perfect sampling an empirical evaluation of the spontaneous magnetization at zero-temperature, finding a non-zero value in $d=3$ and a vanishing value in $d=2$. Next, in Sec. \ref{4}, using a careful coupling with the Bernoulli  site percolation model in $d=2$, we prove rigorously (Theorem \ref{teo41}) that imposing $+$ boundary conditions,  the magnetization in the center of a square box tends to zero in the thermodynamical limit. We further show in Sec. \ref{5}, using again a coupling argument,  that the infinite volume Gibbs measure of the zero-temperature BEG exists and it is unique (Theorems \ref{teo51} and \ref{teo52}).  Finally in  Sec. \ref{6} we prove  the exponential decay of the two-point correlations of the two-dimensional  BEG model at FAD at zero-temperature (Theorem \ref{teo4}). The arguments used in the whole Section \ref{sec:definition} are all  dynamical ones.

Concerning the low temperature  BEG model on
$\mathbb{Z}^2$ at the
AD interface, by a comparison with  a Bernoulli site percolation in a matching graph of the square lattice,  we get  a $(\b,|Y|)$-dependent condition for the vanishing of the infinite volume limit magnetization, improving, for low temperatures, earlier results obtained in \cite{pcl20} via expansion techniques (Theorem \ref{teoAD}).

An anonymous referee of a previous version of this paper  has pointed out to us that the zero-temperature BEG model at the FAD interface coincides with the discrete Widom-Rowlinson model \cite{GL71} with activity value equal to $1$ and in the context of the Widom-Rowlinson model the uniqueness of the Gibbs measure has been established in an old (and quite overlooked) paper  by  Higuchi \cite{Hi83} for a region of activities which includes the value   $1$.

\section{The zero-temperature BEG model at FAD}\label{sec:definition}\mbox{}\\

In what follows, given any finite set $S$, we let $|S|$ be its cardinality.
We will consider $\mathbb{Z}^d$ as the set of vertices (sites) of the  graph $\mathbb{L}^d$ whose edges are the nearest neighbors pairs and given any two points
$x$ and $y$ of $\mathbb{Z}^d$ we let $|x-y|$ be the usual graph distance in $\mathbb{L}^d$. As said in the introduction,
we suppose that a spin variable $\s_x$ taking values in the set $\{0,+1, -1\}$ is sitting in each
site $x\in \Zd$.
Given any finite set $U\subset \Zd$ we will denote by $\Si_U$ the set
of all spin configurations in $U$ (observe that $|\Si_U|=3^{|U|}$).

Hereafter the symbol $\L$ will always denote a finite cubic subset  $\Lambda\subset \mathbb{Z}^d$ with sidelenght $L$ containing in its midpoint the origin $o$ of $\mathbb{Z}^d$ with external and internal boundary given respectively by
$$
\partial_e\L=\{x\in \mathbb{Z}^d\setminus\L: |x-y|=1\;{\rm for\,\, some}\,\, y\in \L\}
$$
and
$$
\partial_i\L=\{x\in \L: |x-y|=1\;{\rm for\,\, some}\,\, y\in \mathbb{Z}^d\setminus\L\}.
$$

From now on, for simplicity, the notation $\lim_{\L\uparrow \infty}$  (i.e. the thermodynamic limit) means  $\lim_{L\to\infty}$. { We stress however that by standard arguments it is possible to prove that the rigorous results obtained in this paper continue to hold  when the limit $\lim_{\L\uparrow \infty}$ is taken along generic increasing sequences of sets invading $\Z^d$}.
A boundary condition $\x$ is a configuration $\x=\{\x_x\}_{x\in \Zd}$
having in mind that, as $L$ tends to $\infty$, the  sites of $\x$ entering in $\L$ are
disregarded and those in  $\Zd\setminus\L$ are kept (so only the boundary configuration $\x\in \Si_{\partial_e\L}$ may influence the bulk).
In particular $\x=+$ is the configuration such that $\x_x=+1$ for all $x\in \Zd$,
$\x=-$ is the configuration such that $\x_x=-1$ for all $x\in \Zd$ and $\x=0$ is the configuration such that $\x_x=0$ for all $x\in \Zd$.

We  define the BEG model  at the FAD interface in  $\Lambda\subset \mathbb{Z}^d$ with boundary conditions  $\x$ via the
 Hamiltonian
\begin{equation}\label{ham}
 H_\Lambda^\x(\sigma)=-\sum_{x\sim y\atop x,y\in \Lambda}\sigma_x\sigma_y(1-\sigma_x\sigma_y)-\sum_{x\in \partial_{i}\Lambda}\,\,\sum_{y\in \partial_e\L, |x-y|=1}(\x_y \sigma_x - \x_y^2\sigma_x^2).
 \end{equation}
Let us denote by $F^\x_\L$  the set of all ground states in $\L$ with fixed boundary conditions, namely
$$
F_\L^\x=\{\s\in \Si_\L: \s_x\s_y\neq -1 ~{\rm for ~all}~ x\sim y \in \L\cup\partial_e\L\}.
$$
Then  the energy $H_\Lambda^\x(\sigma)$ of a configuration $\s\in \Sigma_\L$ is zero whenever $\s\in F^\x_\L$ and it
is positive, being simply twice the number of nearest neighbor edges $\{x,y\}$ such that $\sigma_x\sigma_y=-1$ when
$\s\in\Si_\L\setminus F^\x_\L$. The probability $\PP_{\L,\b}^\x(\s)$ of a given configuration $\s\in \Si_\L$ at finite inverse temperature $\b$  is defined via the Gibbs measure, i.e.
\begin{equation}\label{gibbs}
\PP_{\L,\b}^\x(\s) = {e^{-\b H_\Lambda^\x(\sigma)}\over Z_{\L,\b}^\x}
\end{equation}
where
$$
Z_{\L,\b}^\x = \sum_{\s\in \Si_\L} e^{-\b H_\Lambda^\x(\sigma)}.
$$
When $\beta=\infty$ (i.e. zero temperature) any configuration $\s\in\Si_\L\setminus F^\x_\L$ has zero probability, and hence the finite-volume,
zero-temperature Gibbs measure given in (\ref{gibbs}) becomes the uniform measure on $F^\x_\L$, namely
\begin{equation}\label{gibbs0}
\PP_{\L,\b=\infty}^\x(\s) = \begin{cases}{1\over N_\L^\x} & {\text if}\; \s\in F^\x_\L\\  0 & \text{otherwise}
\end{cases}
\end{equation}
where
$$
N_{\L}^\x = |F^\x_\L| = Z_{\L,\b=+\infty}^\x.
$$
is the number of ground states in $\L$ with boundary condition $\x$ outside $\L$.
In the rest of   this section   we will omit the index $\b=\infty$ for the  zero temperature Gibbs measure with $\x$ boundary conditions and denote it with
the shorten symbol  $\PP_{\L}^\x(\s)$.
Given a function $f:\Si_\L\to \mathbb{R}$ we will denote by $\<f(\s)\>_{\L}^\x$
the expected value of $f(\s)$ with respect to the uniform  measure $\PP_{\L}^\x(\s)$   under $\x$ boundary conditions.

\subsection{Sampler for the BEG model at FAD interface, zero temperature}\label{3}\mbox{}\\

In the present section we will perform  a numerical analysis of the expected value of the spin at the origin under $+$ boundary conditions, i.e.
the following quantity.
\begin{equation}\label{mag}
\<\s_o\>_\L^+={1\over N^+_\L}\sum_{\s\in F^+_\L}\s_o
={1\over N^+_\L}\left[\sum_{\sigma\in{F^+_{\Lambda}}\atop\s_o=+1}1-\sum_{\s\in F^+_\L\atop\s_o=-1}1\right].
\end{equation}

We will further study numerically  the two-point correlation  function  under free  boundary conditions, namely
\be\label{corre}
\< \s_{x}\s_y\>_\L^0=
{1\over   N^0_\L}\left[\sum_{\sigma\in{F^0_{\Lambda}}\atop\s_x\s_y=+1}1-\sum_{\s\in F^0_\L\atop\s_x\s_y=-1}1\right].
\ee

 In order to do that we will define a symmetric and ergodic Markov Chain whose stationary measure
coincides with the zero-temperature Gibbs measure  of the BEG model at FAD. We also will introduce a coupling between two Markov chains as above on systems with  different boundary conditions which preserve the  natural partial order in the configuration set
$\{-1,+1,0\}^\L$. This coupling allows us to perform a perfect sampling simulation of our system
producing numerical results on
the behavior of $\<\s_o\>_\L^+$
in two and three dimensions as the size of the box $\L$  increases, and on the decay of the two-point correlations
$\< \s_{x}\s_y\>_\L^0$  in $d=2$ for different values of the distance $|x-y|$.

\vskip.2cm

\\{\bf The Markov Chain}. Given a  box $\L$ and a boundary condition $\x$ (hence spins at sites of $\partial_e\L$ are fixed), we recall that
 the Gibbs distribution at zero temperature is uniform on $F^\x_\L$. We call feasible a configuration $\s\in F^\x_\L$ and  we define a Markov chain that is symmetric ($P_{\sigma\tau}=P_{\tau\sigma}$) for all pairs of feasible configurations $\sigma,\tau$, while $P_{\sigma\tau}=0$ when at least  one between $\sigma$ and $\tau$ is not feasible. More explicitly, the Markov chain is defined as follows: assume $\sigma$ feasible, and call $N_x(\sigma)$ the set of values of the spin that are present in the neighborhood of the site $x$. The transition probabilities of our sampler are defined to be $P_{\sigma\tau}=0$ if $\sigma$ and $\tau$ differ in more than one site while, for couple of configurations differing at most in one site, they are defined by the following procedure.

\begin{enumerate}
\item Choose a site $x\in\L$ uniformly at random (u.a.r.),

\item Set $\tau_y=\sigma_y$ for all $y\ne x$.

 \item Set in the configuration $\tau$ the value $\tau_x$ of the spin in the site $x$ with uniform distribution among the feasible values.  This means that
\end{enumerate}

\begin{itemize}
\item if $N_x(\sigma)=\{0\}$ then $\tau_x=1,0,-1$ with probability $1/3$;

\item if $N_x(\sigma)=\{0, +1\}$ or $N_x(\sigma)=\{ +1\}$ then
$\tau_x=0,+1$ with probability $1/2$;

\item if $N_x(\sigma)=\{0, -1\}$ or $N_x(\sigma)=\{ -1\}$ then
$\tau_x=0,-1$ with probability $1/2$;

\item  if $N_x(\sigma)$ contains both $-1$ and $+1$ then
$\tau_x=0$ with probability $1$.
\end{itemize}

This procedure defines a Markov chain $X(t)$ ($t=0,1,2,\dots$) with the following features. First of all, the chain is ergodic. To prove this it is sufficient to
observe that in a finite number of steps it is possible to reach with nonzero probability any state starting from any state: simply, pass through the state $\sigma_x=0$ for all $x\in\Lambda$. The chain is obviously aperiodic, since $P_{\sigma\sigma}\ne 0$ for all $\sigma$.
Second, starting from a feasible configuration, the evolution of the system remains on feasible configurations, since according to the rules above it is impossible to create an edge $x\sim y$ such that $\sigma_x\sigma_y=-1$.

Third, the probability transition matrix $P_{\sigma\tau}$  is evidently symmetric, $P_{\sigma\tau}=P_{\t\s}$.  Hence the (unique)
stationary measure of our chain is uniform on all the feasible configurations, and therefore it coincides
with  $\PP^\x_\L$  (i.e. the uniform Gibbs measure of the  zero-temperature  BEG model at FAD).

Observe now that the spin configurations $\s\in F^\x_\L$ are partially ordered:
the partial order relation  $\sigma\prec\tau$ is defined trivially by $\sigma\prec\tau\Leftrightarrow \sigma_x\le\tau_x$, $\forall x\in\Lambda$. This circumstance permits
to define an order preserving coupling in the implementation of our Markov chain. Namely, we let evolve together two chains $X$ and $X'$ (starting in general from two different initial spin configurations  $\s(0)$ and $\s'(0)$)  in such a way that the marginal of the
evolution of each chain represents exactly the probabilities $P_{\s\t}$ defined above, but the two evolutions
are coupled: if the initial configurations $\s(0)$ and $\s'(0)$  are such that $\s(0)\prec\s'(0)$, then
their evolutions $\s(t)$ and $\s'(t)$ are such that $\s(t)\prec\s'(t)$ at any $t\ge 1$. This is easily realized by defining judiciously the updating rules of the two
chains accordingly to the values of the sets $N_x(\s(t))$ and $N_x(\s'(t))$. Assuming thus that at step $0$ we have  $\s(0)\prec \s'(0)$, our
coupling will be defined as follows. At each step $t$ of the implementation of the Markov chain we
update in the two configurations $\s(t)$ and $\s'(t)$ by chosing u.a.r. a site $x\in \L$, extracting then a single random variable $U$
uniformly distributed in $[0, 1]$ and  letting $\s(t+1)$ and $\s'(t+1)$ such that  $\s_y(t+1)=\s_y(t)$ and
$\s'_y(t+1)=\s'_y(t)$  for all $y\in \L\setminus\{x\}$
and setting  $\s_x(t+1)$ and $\s'_x(t+1)$  according to the
\emph{same} value of U as follows. We define a set of thresholds in the segment $[0, 1]$ according to the
sets $N_x(\s(t))$ and $N_x(\s'(t))$  exploiting the fact that the probability of $U$ is uniform and hence $U$ enters
a segment of length $\ell$ contained in  $[0,1]$  with probability $\ell$. The thresholds however are fixed in such a way that we
will always have $\s_x(t+1)\le\s'_x(t+1)$. We list in the Table 1 all possible pairs  $N_x(\s(t))$ and $N_x(\s'(t))$  together with the
drawings of the segments $[0, 1]$ with relative thresholds for $\s_x(t+1)$ and $\s'_x(t+1)$.

\begin{table}[H]
\renewcommand{\arraystretch}{1.2}
\begin{center}
\begin{tabular}{|c||c||c|}\hline
$N_x{\big(\sigma(t)\big)}$ & $N_x{\big(\s'(t)\big)}$ & Thresholds \\
\hline\hline
$\{0,-1\}$ or $\{-1\}$ &
$\{0\}$ &
\begin{tikzpicture}[baseline=-0.7ex,scale=0.5]
      \draw [thick] (-3,0.5) -- (3,0.5);
      \node [right] at (3.1,0.5) {\tiny{$\s_x(t+1)$}};
      \draw (-3,0.3) -- (-3,0.7);
      \draw (0,0.3) -- (0,0.7);
      \draw (3,0.3) -- (3,0.7);
      \node [above] at (-1.5,0.6){\tiny{$-1$}};
      \node [above] at (1.5,0.6) {\tiny{$0$}};

      \draw [thick] (-3,-0.5) -- (3,-0.5);
      \node [right] at (3.1,-0.5) {\tiny{$\s'_x(t+1)$}};
      \draw (-3,-0.7) -- (-3,-0.3);
      \draw (-1,-0.7) -- (-1,-0.3);
      \draw (1,-0.7) -- (1,-0.3);
      \draw (3,-0.7) -- (3,-0.3);
      \node [above] at (-2,-0.4){\tiny{$-1$}};
      \node [above] at (0,-0.4) {\tiny{$0$}};
      \node [above] at (2,-0.4) {\tiny{$1$}};
      \end{tikzpicture} \\
\hline
$\{0,-1\}$ or $\{-1\}$  & $\{0,-1\}$ or $\{-1\}$ &
      \begin{tikzpicture}[baseline=-0.7ex,scale=0.5]
      \draw [thick] (-3,0.5) -- (3,0.5);
      \node [right] at (3.1,0.5) {\tiny{$\s_x(t+1)$}};
      \draw (-3,0.3) -- (-3,0.7);
      \draw (0,0.3) -- (0,0.7);
      \draw (3,0.3) -- (3,0.7);
      \node [above] at (-1.5,0.6){\tiny{$-1$}};
      \node [above] at (1.5,0.6) {\tiny{$0$}};

      \draw [thick] (-3,-0.5) -- (3,-0.5);
      \node [right] at (3.1,-0.5) {\tiny{$\s'_x(t+1)$}};
      \draw (-3,-0.7) -- (-3,-0.3);
      \draw (0,-0.7) -- (0,-0.3);
      \draw (3,-0.7) -- (3,-0.3);
      \node [above] at (-1.5,-0.4){\tiny{$-1$}};
      \node [above] at (1.5,-0.4) {\tiny{$0$}};
      \end{tikzpicture}\\
\hline
 $\{0,-1\}$ or $\{-1\}$  & $\{0,+1\}$ or \{+1\} &
      \begin{tikzpicture}[baseline=-0.7ex,scale=0.5]
      \draw [thick] (-3,0.5) -- (3,0.5);
      \node [right] at (3.1,0.5) {\tiny{$\sigma_x(t+1)$}};
      \draw (-3,0.3) -- (-3,0.7);
      \draw (0,0.3) -- (0,0.7);
      \draw (3,0.3) -- (3,0.7);
      \node [above] at (-1.5,0.6){\tiny{$-1$}};
      \node [above] at (1.5,0.6) {\tiny{$0$}};

      \draw [thick] (-3,-0.5) -- (3,-0.5);
      \node [right] at (3.1,-0.5) {\tiny{$\s'_x(t+1)$}};
      \draw (-3,-0.7) -- (-3,-0.3);
      \draw (0,-0.7) -- (0,-0.3);
      \draw (3,-0.7) -- (3,-0.3);
      \node [above] at (-1.5,-0.4){\tiny{$+1$}};
      \node [above] at (1.5,-0.4) {\tiny{$0$}};
      \end{tikzpicture}\\
\hline
$\{0,-1\}$ or $\{-1\}$  & $\{0,\pm1\}$ &
      \begin{tikzpicture}[baseline=-0.7ex,scale=0.5]
      \draw [thick] (-3,0.5) -- (3,0.5);
      \node [right] at (3.1,0.5) {\tiny{$\sigma_x(t+1)$}};
      \draw (-3,0.3) -- (-3,0.7);
      \draw (0,0.3) -- (0,0.7);
      \draw (3,0.3) -- (3,0.7);
      \node [above] at (-1.5,0.6){\tiny{$-1$}};
      \node [above] at (1.5,0.6) {\tiny{$0$}};

      \draw [thick] (-3,-0.5) -- (3,-0.5);
      \node [right] at (3.1,-0.5) {\tiny{$\s'_x(t+1)$}};
      \draw (-3,-0.7) -- (-3,-0.3);
      \draw (3,-0.7) -- (3,-0.3);
      \node [above] at (0,-0.4) {\tiny{$0$}};
      \end{tikzpicture}
    \\
    \hline\hline
    $\{0\}$ & $\{0\}$ &
      \begin{tikzpicture}[baseline=-0.7ex,scale=0.5]
      \draw [thick] (-3,0.5) -- (3,0.5);
      \node [right] at (3.1,0.5) {\tiny{$\sigma_x(t+1)$}};
      \draw (-3,0.3) -- (-3,0.7);
      \draw (-1,0.3) -- (-1,0.7);
      \draw (1,0.3) -- (1,0.7);
      \draw (3,0.3) -- (3,0.7);
      \node [above] at (-2,0.6){\tiny{$-1$}};
      \node [above] at (0,0.6) {\tiny{$0$}};
      \node [above] at (2,0.6) {\tiny{$1$}};

      \draw [thick] (-3,-0.5) -- (3,-0.5);
      \node [right] at (3.1,-0.5) {\tiny{$\s'_x(t+1)$}};
      \draw (-3,-0.7) -- (-3,-0.3);
      \draw (-1,-0.7) -- (-1,-0.3);
      \draw (1,-0.7) -- (1,-0.3);
      \draw (3,-0.7) -- (3,-0.3);
      \node [above] at (-2,-0.4){\tiny{$-1$}};
      \node [above] at (0,-0.4) {\tiny{$0$}};
      \node [above] at (2,-0.4) {\tiny{$1$}};
      \end{tikzpicture}
    \\
\hline
$\{0\}$ & $\{0,+1\}$ or $\{1\}$ &
      \begin{tikzpicture}[baseline=-0.7ex,scale=0.5]
      \draw [thick] (-3,0.5) -- (3,0.5);
      \node [right] at (3.1,0.5) {\tiny{$\sigma_x(t+1)$}};
      \draw (-3,0.3) -- (-3,0.7);
      \draw (-1,0.3) -- (-1,0.7);
      \draw (1,0.3) -- (1,0.7);
      \draw (3,0.3) -- (3,0.7);
      \node [above] at (-2,0.6){\tiny{$-1$}};
      \node [above] at (0,0.6) {\tiny{$0$}};
      \node [above] at (2,0.6) {\tiny{$1$}};

      \draw [thick] (-3,-0.5) -- (3,-0.5);
      \node [right] at (3.1,-0.5) {\tiny{$\s'_x(t+1)$}};
      \draw (-3,-0.7) -- (-3,-0.3);
      \draw (0,-0.7) -- (0,-0.3);
      \draw (3,-0.7) -- (3,-0.3);
      \node [above] at (-1.5,-0.4){\tiny{$0$}};
      \node [above] at (1.5,-0.4) {\tiny{$1$}};
      \end{tikzpicture}
    \\
\hline\hline
    $\{0,\pm1\}$ & $\{0,+1\}$ or $\{1\}$ &
      \begin{tikzpicture}[baseline=-0.7ex,scale=0.5]
      \draw [thick] (-3,0.5) -- (3,0.5);
      \node [right] at (3.1,0.5) {\tiny{$\sigma_x(t+1)$}};
      \draw (-3,0.3) -- (-3,0.7);
      \draw (3,0.3) -- (3,0.7);
      \node [above] at (0,0.6) {\tiny{$0$}};

      \draw [thick] (-3,-0.5) -- (3,-0.5);
      \node [right] at (3.1,-0.5) {\tiny{$\s'_x(t+1)$}};
      \draw (-3,-0.7) -- (-3,-0.3);
      \draw (0,-0.7) -- (0,-0.3);
      \draw (3,-0.7) -- (3,-0.3);
      \node [above] at (-1.5,-0.4){\tiny{$0$}};
      \node [above] at (1.5,-0.4) {\tiny{$1$}};
      \end{tikzpicture}
    \\
\hline
$\{0,\pm1\}$ & $\{0,\pm1\}$ &
      \begin{tikzpicture}[baseline=-0.7ex,scale=0.5]
      \draw [thick] (-3,0.5) -- (3,0.5);
      \node [right] at (3.1,0.5) {\tiny{$\sigma_x(t+1)$}};
      \draw (-3,0.3) -- (-3,0.7);
      \draw (3,0.3) -- (3,0.7);
      \node [above] at (0,0.6) {\tiny{$0$}};

      \draw [thick] (-3,-0.5) -- (3,-0.5);
      \node [right] at (3.1,-0.5) {\tiny{$\s'_x(t+1)$}};
      \draw (-3,-0.7) -- (-3,-0.3);
      \draw (3,-0.7) -- (3,-0.3);
      \node [above] at (0,-0.4) {\tiny{$0$}};
      \end{tikzpicture}\\
\hline\hline
    $\{0,+1\}$ or $\{1\}$ & $\{0,+1\}$ or $\{1\}$ &
      \begin{tikzpicture}[baseline=-0.7ex,scale=0.5]
      \draw [thick] (-3,0.5) -- (3,0.5);
      \node [right] at (3.1,0.5) {\tiny{$\sigma_x(t+1)$}};
      \draw (-3,0.3) -- (-3,0.7);
      \draw (0,0.3) -- (0,0.7);
      \draw (3,0.3) -- (3,0.7);
      \node [above] at (-1.5,0.6){\tiny{$0$}};
      \node [above] at (1.5,0.6) {\tiny{$1$}};

      \draw [thick] (-3,-0.5) -- (3,-0.5);
      \node [right] at (3.1,-0.5) {\tiny{$\s'_x(t+1)$}};
      \draw (-3,-0.7) -- (-3,-0.3);
      \draw (0,-0.7) -- (0,-0.3);
      \draw (3,-0.7) -- (3,-0.3);
      \node [above] at (-1.5,-0.4){\tiny{$0$}};
      \node [above] at (1.5,-0.4) {\tiny{$1$}};
      \end{tikzpicture}\\
\hline
\end{tabular}
\end{center}
\caption{Order preserving coupling}
\label{ta1}
\end{table}

Aiming to perform  computer simulations using this Markov chain, an important feature of the order preserving coupling  described above is that it is possible to perform with it  a perfect sampling on the stationary measure.
We choose the boundary condition $\x=+$ and let   $\s^{\rm min}\in F^+_\L$ be the configuration such that
$\s^{\rm min}_x=-1$ for all $x\in \L\setminus\partial_i\L$ and
$\s^{\rm min}_x=0$ for $x\in \partial_i\L$. Moreover we let  $\s^{\rm max}\in F^+_\L$ be the configuration such that
 $\s^{\rm max}_x=+1$ for all $x\in \L$. Clearly  $\s^{\rm min}$  and $\s^{\rm max}$ are  the infimum and the supremum respectively over all configurations in $F^+_\L$, i.e. for  all the configurations $\sigma\in F^+_\L$
 we have $\s^{\rm min}\prec\sigma\prec \s^{\rm max}$.

We can now run the coupled chains, starting respectively from  $\s(0)=\s^{\rm min}$ and $\s'(0)=\s^{\rm max}$
according to the standard procedure of perfect sampling (coupling from the past). It is not difficult to prove (see for instance  \cite{hagg02}  for an introductory reference) that the obtained configurations are distributed uniformly, i.e. accordingly with the stationary measure.
We computed empirically, according to this procedure,  the magnetization in the origin for various value of $|\Lambda|$ in $2$ and $3$ dimensions.
\begin{figure}
\begin{minipage}[c]{0.5\textwidth}
\includegraphics[width=\textwidth]{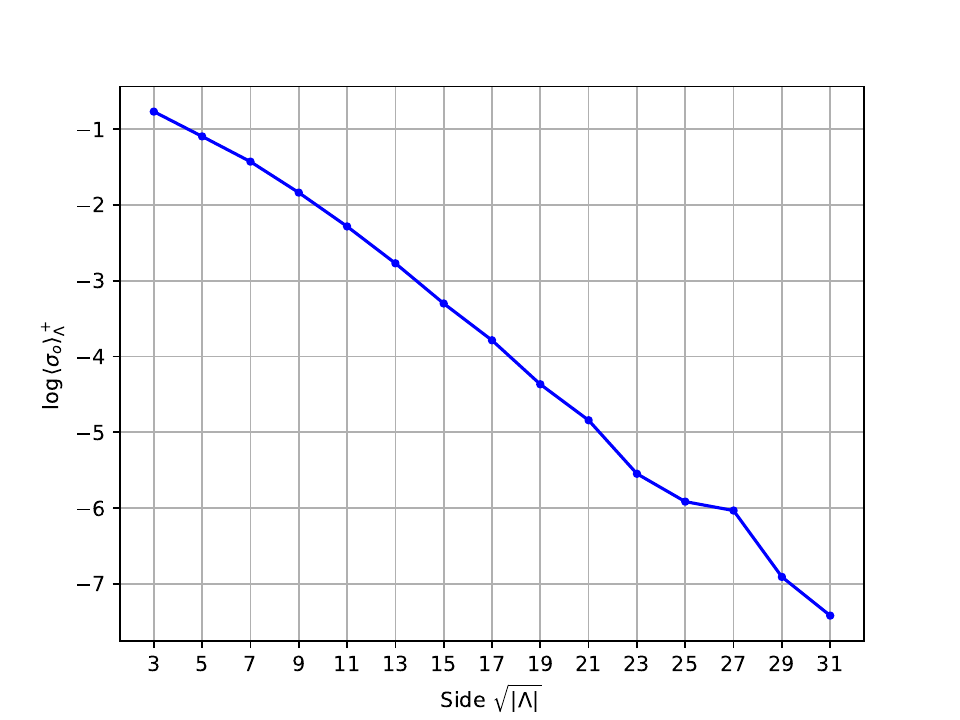}
\end{minipage}
\hfill
\begin{minipage}[c]{0.5\textwidth}
\includegraphics[width=\textwidth]{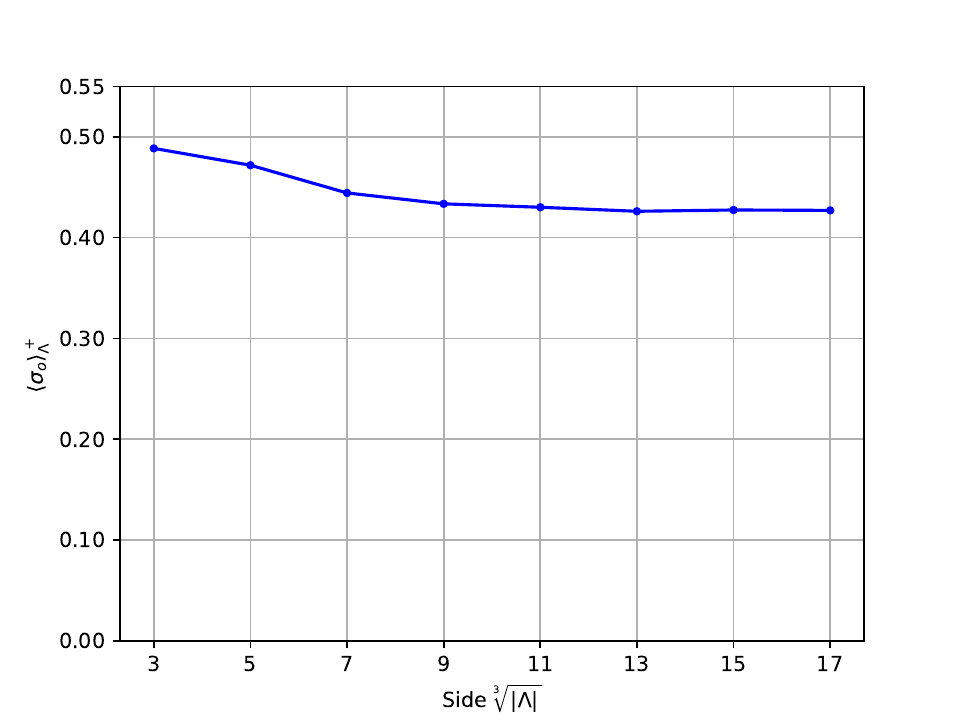}
\end{minipage}
\caption{\label{fig:mag_d2}Magnetization at the origin $o$ for the zero-temperature BEG model in $d=2$ (left) and $d=3$ (right). Results of $10^6$ samples obtained running the BEG model sampler according to the perfect sampling procedure.}
\end{figure}

The results we obtained show quite clearly that in 2 dimensions the magnetization in the origin tends to vanish, while in 3 dimensions it tends to a strictly positive value. Figure \ref{fig:mag_d2}  clarifies the previous statement.

\begin{figure}[H]
  \centering
  \includegraphics[width=10cm]{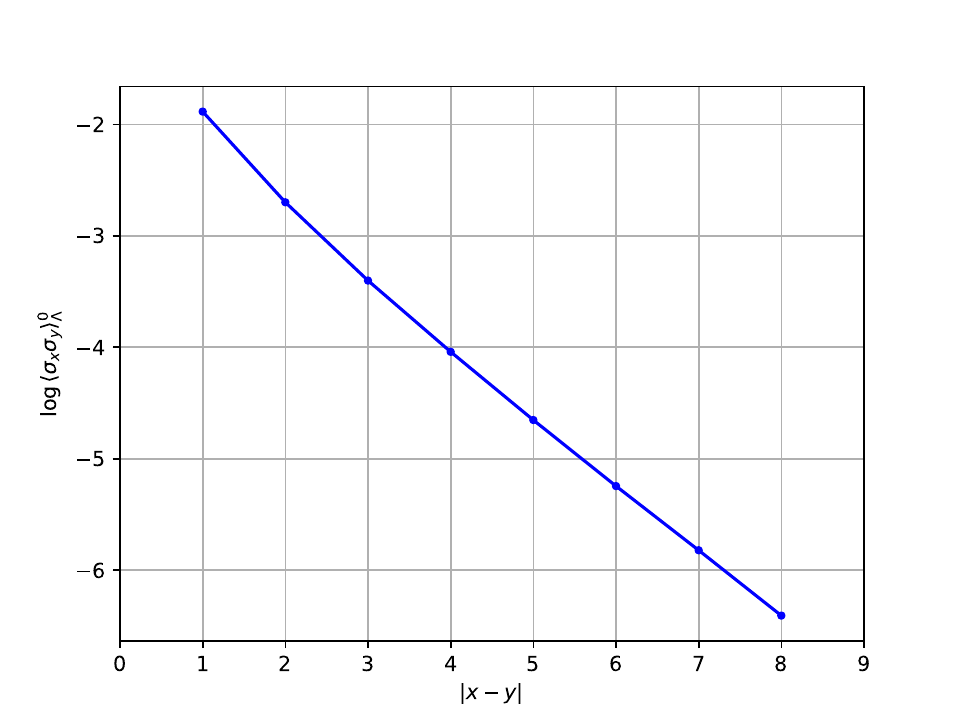}
  \caption{Two-point correlation for the BEG model in $d=2$. Results of $2\cdot10^6$ samples obtained running the BEG model sampler according to the perfect sampling procedure in a box $\Lambda$ of side $17$. Correlations are thus computed at different spin distances $|x-y|$.}
\label{fig:2p_correlation}
\end{figure}

Furthermore, we applied the BEG sampler to compute empirically the two-point correlations in 2-dimensions at various distances $|x-y| \in \{ 1,2,\dots,8\}$ around the origin $o$.
Figure \ref{fig:2p_correlation} shows the results on logarithmic scale where it appears clearly the exponential nature of the decay of the two-point correlations.

The rest of the section  is  devoted to a rigorous study of the  two-dimensional, zero-temperature BEG model at FAD.
In this study the Markov chain and the coupling introduced defined above  will play a fundamental role.

\subsection{Magnetization at zero temperature in the $d=2$ BEG model at FAD}\label{4}\mbox{}\\

Here our goal  is to analyze  rigorously the behaviour of
the expected value $\<\s_o\>_{\L}^+$ of the spin  at  the origin $o$ defined in (\ref{mag}) as $\L\uparrow\infty$ and when $d=2$.
The thermodynamic limit of $\<\s_o\>_{\L}^+$, if it exists,  will be denoted hereafter by $\langle\sigma_o\rangle^+$, i.e.
\begin{equation}
\langle\sigma_o\rangle^+=\lim_{\Lambda\uparrow\infty}\langle\sigma_o\rangle_{\Lambda}^+.
\end{equation}

The main result of this  section is the following theorem.

\begin{theorem}\label{teo41}
For the BEG model at zero temperature and at FAD interface in $d=2$ we have that
$$
\langle\sigma_0\rangle^+=\langle\sigma_0\rangle^-=0.
$$
\end{theorem}

In order to prove Theorem \ref{teo41} we start by proving
the following lemma.
\begin{lemma}\label{lem42}
The expected value of the spin at the origin  for the two-dimensional and zero-temperature BEG model at FAD  confined in the box $\L$ with $+$ boundary conditions defined in (\ref{mag}) admits the bound
\begin{equation}\label{mag2}
\langle\sigma_0\rangle^+_{\Lambda}
=
\PP_\L^+(o\rightsquigarrow\partial\L)
\end{equation}
where $\PP_\L^+$ is the zero temperature Gibbs measure with $+$ boundary conditions defined in (\ref{gibbs0}) and the symbol $o\rightsquigarrow\partial\L$ denotes the event that
the origin $o$ is connected to a point $z\in \partial_e\L$ through a path $p$ such that $\s_x\neq 0$ for all $x\in p$.
\end{lemma}

\noindent{\bf Proof}.  Firt of all note that we  can  rewrite the quantity $\<\s_0\>^+_\L$ defined in (\ref{mag}) as
\begin{equation}\label{split}
\<\s_o\>^+_\L={1\over N^+_\L}\left[  N^+_\L|_{\s_o=+1} -   N^+_\L|_{\s_o=-1}\right]
\end{equation}
where $N^+_\L|_{\s_o=\pm1}$ is the number of ground states in $\L$ with $+$ boundary conditions at $\partial_e\L$ and
with the spin at the origin fixed at the value $\s_o=\pm 1$. We further denote by $o\not\rightsquigarrow\partial_e\L$
the complementary event of  $o\rightsquigarrow\partial_e\L$ and by
$N^+_\L|_{\s_o=\pm 1, o\rightsquigarrow\partial\L }$ (resp. $N^+_\L|_{\s_o=\pm1, o\not\rightsquigarrow\partial\L }$)
the number of ground states in $\L$ with the spin at the origin fixed at the value $\s_o=\pm 1$ and such that
 $o\rightsquigarrow\partial\L$ (resp. $o\not\rightsquigarrow\partial\L$) and we  observe that
\be\label{N+}
 N^+_\L|_{\s_o=+1}=  N^+_\L|_{\s_o=+1, o\rightsquigarrow\partial\L }+
 N^+_\L|_{\s_o=+1, o\not\rightsquigarrow\partial\L }.
\ee
while
\be\label{N-}
N^+_\L|_{\s_o=-1}=   N^+_\L|_{\s_o=-1, o\not\rightsquigarrow\partial\L }
\ee
since clearly $N^+_\L|_{\s_o=-1, o\rightsquigarrow\partial\L }=0$.  Inserting (\ref{N+}) and (\ref{N-}) into
(\ref{split}) we get
$$
\<\s_0\>^+_\L={1\over N^+_\L}\left[ N^+_\L|_{\s_o=+1, o\rightsquigarrow\partial\L }+N^+_\L|_{\s_o=+1, o\not\rightsquigarrow\partial\L } -   N^+_\L|_{\s_o=-1, o\not\rightsquigarrow\partial\L }\right]
$$
Let us  introduce some further  notations.
A set of vertices $\g\subset\L$ is said to a {\it barrier} (surrounding the origin) if any path starting at the origin and ending at some vertex of $\partial_e\L$ contains a vertex of $\g$. A  {\it  contour} (surrounding the origin)  in $\L$ is a barrier $\g\subset \L$ such that, for any $x\in \g$,  $\g\setminus\{x\}$ is not a barrier. We denote by $\CC^o_\L$ the set of all  contours surrounding the origin  in $\L$. Given $\g\in \CC^o_\L$, the interior of $\g$, denoted by $I_\g$, is the set formed by those  vertices $y\in\L$ for which there exists a path $p$ starting at  the origin $o$  and ending at $y$
such that $p\cap \g=\0$. We also denote $E^\L_\g=\L\setminus (\g\cup I_\g)$.  Finally given a  contour $\g$ and a configuration $\s\in \Si_\L$, we set
$I_\g^0(\s)= \{x\in I_{\g}: \s_x=0\}$.
Now note that  if $\s$ is a configuration in $\L$ belonging to the event  $o\not\rightsquigarrow\partial_e\L$, then necessarily there is a unique
minimal contour $\g_\s$ such that $\s_x=0$ for all $x\in \g_\s$ and such that  the unique   contour contained in the
subset $I^0_{\g_\s}(\s)\cup \g_\s$   is $\g_\s$.
Given $\g\in \CC^0_\L$, let us denote by $N^+_\L|_{\s_0=\pm 1, \g}$ the number of ground states $\s\in F^+_\L$ such that
$\s_o=\pm1$, $o\not\rightsquigarrow\partial_e\L$ and $\g_\s=\g$. Then we have that
$$
N^+_\L|_{\s_0=\pm 1, \g}=N^{+}(E^\L_\g)N^*_{\s_o=\pm1}(I_\g)
$$
where  $N^{+}(E^\L_\g)$ is the number of ground states in $E^\L_\g$ with boundary conditions $+$ at $\partial_e\L$ and
zero at $\g$ and $N^*_{\s_o=\pm1}(I_\g)$ is the number of ground states $\s$ in $\g\cup I_\g$ (with zero boundary condition at $\g$) under the constraint that $\s_0=\pm1$ and that the unique   contour contained in $\g\cup I^0_\g(\s)$ is   $\g$.
With these notations we can write
$$
N^+_\L|_{\s_o=\pm1, o\not\rightsquigarrow\partial\L }= \sum_{\g\in \CC^0_\L}N^+_\L|_{\s_0=\pm 1, \g}=
\sum_{\g\in \CC^0_\L}N^+(E^\L_\g)N^*_{\s_0=\pm 1}(I_\g)
$$
Therefore we have that
\be\label{sigo}
 \<\s_0\>^+_\L(\b=+\infty)={1\over N^+(\L)}\left[ N^+_{\L}|_{\s_o=+1,\,0^{+}_\rightsquigarrow\partial\L}+ \sum_{\g\in \CC^0_\L} N^+(E^\L_\g)\Big[N^*_{\s_0=+1}(I_\g)-N^*_{\s_0=-1}(I_\g)\Big]\right]
\ee
But now, by the spin flip symmetry we have that, for any $\g\in \CC^o_\L$
\be\label{symme}
N^*_{\s_0=+1}(I_\g)=N^*_{\s_0=-1}(I_\g)
\ee
and so, inserting (\ref{symme}) into (\ref{sigo}) we get
\be\label{true}
\<\s_0\>^+_\L={N^+_\L|_{\s_o=+1, o\rightsquigarrow\partial\L }\over N^+_\L}= \PP_\L^+(o\rightsquigarrow\partial\L ).
\ee
which concludes the proof of Lemma \ref{lem42}. \qed

\vglue.2truecm

Lets us now consider the Bernoulli site percolation in $\Z^2$ with parameter $p={1\over 2}$.
Namely, suppose that in each site $x\in \Z^2$ an independent variable $\om_x$ is defined. Such $\om_x$ takes the value $\om_x=+1$ with probability ${1\over 2}$ indicating that the site is open.
When $\om_x$ takes the value $\om_x=0$ with probability ${1\over 2}$ it indicates that the  site is closed.
Given a square  $\L\subset \Z^2$ centered at the origin,  let $\O_\L=\{0,1\}^\L $ be the set of all possible site configurations in $\L$. We will denote by  $\PP^{\rm perc}_\L$ the probability product measure of the site percolation on $\O_\L=\{0,1\}^\L $  for $p={1\over 2}$.

Let us introduce a Markov chain $Y(t)$  on $\O_\L$ whose  transition probabilities $P_{\om\om'}$ for $\om, \om'\in \O_\L$ are defined by the following sampler.

\begin{enumerate}
\item Choose a site $x\in \L$ uniformly at random.

\item Set $\om'_y=\om_y$ for all $y\in \L\setminus\{x\}$.

\item  Extract $U$ uniformly distributed in $[0,1]$, and set  $\om'_x=+1$ if $U<1/2$, and set
$\om'_x=0$, otherwise.
\end{enumerate}

It is immediate to see that this is a sampler of the Bernoulli site percolation with $p=1/2$, and that $Y(t)$ is distributed exactly according
the probability measure $\PP^{\rm perc}_\L$   provided at time $t$ each site $x$ has been visited at least once.

\mbox{}

 \noindent{\bf Percolation-BEG coupling}. Given a box $\L\subset \Z^2$, let $Y(t)$ be the Markov chain described above with  stationary measure $\PP^{\rm perc}_\L$ and let
$X(t)$ be the Markov chain introduced  in Section \ref{3} with stationary measure $\PP^+_\L$ (i.e. the uniform Gibbs measure of the  zero-temperature  BEG model at FAD with $+$ boundary conditions). We
define a coupling between $Y(t)$ and $X(t)$ as follows.
Assume the initial site configuration  $Y(0)$ is the configuration where all sites are closed (i.e. $\om_x(0)=0$ for all $x\in \L$) and the initial spin configuration $X(0)$ is the configuration where all spins are zero (i.e.  $\s_x(0)=0$ for all $x\in \L$).
We then let evolve together the two chains $Y$ and $X$  in such a way that at each step $t$ of the implementation of the two coupled Markov Chains we update  the two configurations $\om(t)$ and $\s(t)$ by choosing u.a.r. a site $x\in \L$, extracting then a single random variable $U$
uniformly distributed in $[0, 1]$ and  letting $\om(t+1)$ and $\s(t+1)$ be such that  $\om_y(t+1)=\om_y(t)$ and
$\s_y(t+1)=\s_y(t)$   for all $y\in \L\setminus\{x\}$
and setting  $\s_x(t+1)$ and $\om_x(t+1)$  according to the
same value of U as illustrated in the following table.

\begin{table}[H]
\renewcommand{\arraystretch}{1.2}
\begin{center}
\begin{tabular}{|c||c||c|}\hline
$N_x{\big(\om(t)\big)}$ & $N_x{\big(\s(t)\big)}$ & Thresholds \\
\hline\hline
any &
$\{0,-1\}$ or $\{-1\}$  &
\begin{tikzpicture}[baseline=-0.7ex,scale=0.5]
      \draw [thick] (-3,0.5) -- (3,0.5);
      \node [right] at (3.1,0.5) {\tiny{$\om_x(t+1)$}};
      \draw (-3,0.3) -- (-3,0.7);
      \draw (0,0.3) -- (0,0.7);
      \draw (3,0.3) -- (3,0.7);
      \node [above] at (-1.5,0.6){\tiny{$+1$}};
      \node [above] at (1.5,0.6) {\tiny{$0$}};
      \draw [thick] (-3,-0.5) -- (3,-0.5);
      \node [right] at (3.1,-0.5) {\tiny{$\s_x(t+1)$}};
      \draw (-3,-0.7) -- (-3,-0.3);
      \draw (0,-0.7) -- (0,-0.3);
      \draw (3,-0.7) -- (3,-0.3);
      \node [above] at (-1.5,-0.4){\tiny{$-1$}};
      \node [above] at (1.5,-0.4) {\tiny{$0$}};
\end{tikzpicture}
    \\
    \hline\hline
     any & $\{0\}$ &
      \begin{tikzpicture}[baseline=-0.7ex,scale=0.5]
      \draw [thick] (-3,0.5) -- (3,0.5);
      \node [right] at (3.1,0.5) {\tiny{$\om_x(t+1)$}};
      \draw (-3,0.3) -- (-3,0.7);
      \draw (0,0.3) -- (0,0.7);
      \draw (3,0.3) -- (3,0.7);
      \node [above] at (-1.5,0.6){\tiny{$+1$}};
      \node [above] at (1.5,0.6) {\tiny{$0$}};

      \draw [thick] (-3,-0.5) -- (3,-0.5);
      \node [right] at (3.1,-0.5) {\tiny{$\s_x(t+1)$}};
      \draw (-3,-0.7) -- (-3,-0.3);
      \draw (-1,-0.7) -- (-1,-0.3);
      \draw (1,-0.7) -- (1,-0.3);
      \draw (3,-0.7) -- (3,-0.3);
      \node [above] at (-2,-0.4){\tiny{$+1$}};
      \node [above] at (0,-0.4) {\tiny{$0$}};
      \node [above] at (2,-0.4) {\tiny{$-1$}};
      \end{tikzpicture}
    \\
\hline\hline
   any  & $\{0,+1\}$ or $\{+1\}$ &
      \begin{tikzpicture}[baseline=-0.7ex,scale=0.5]
      \draw [thick] (-3,0.5) -- (3,0.5);
      \node [right] at (3.1,0.5) {\tiny{$\om_x(t+1)$}};
      \draw (-3,0.3) -- (-3,0.7);
      \draw (0,0.3) -- (0,0.7);
      \draw (3,0.3) -- (3,0.7);
      \node [above] at (-1.5,0.6){\tiny{$+1$}};
      \node [above] at (1.5,0.6) {\tiny{$0$}};

      \draw [thick] (-3,-0.5) -- (3,-0.5);
      \node [right] at (3.1,-0.5) {\tiny{$\s_x(t+1)$}};
      \draw (-3,-0.7) -- (-3,-0.3);
      \draw (0,-0.7) -- (0,-0.3);
      \draw (3,-0.7) -- (3,-0.3);
      \node [above] at (-1.5,-0.4){\tiny{$+1$}};
      \node [above] at (1.5,-0.4) {\tiny{$0$}};
      \end{tikzpicture}
    \\
\hline
any& $\{0,\pm1\}$ &
      \begin{tikzpicture}[baseline=-0.7ex,scale=0.5]
      \draw [thick] (-3,0.5) -- (3,0.5);
      \node [right] at (3.1,0.5) {\tiny{$\om_x(t+1)$}};
      \draw (-3,0.3) -- (-3,0.7);
      \draw (0,0.3) -- (0,0.7);
      \draw (3,0.3) -- (3,0.7);
      \node [above] at (-1.5,0.6){\tiny{$+1$}};
      \node [above] at (1.5,0.6) {\tiny{$0$}};

      \draw [thick] (-3,-0.5) -- (3,-0.5);
      \node [right] at (3.1,-0.5) {\tiny{$\s_x(t+1)$}};
      \draw (-3,-0.7) -- (-3,-0.3);
      \draw (3,-0.7) -- (3,-0.3);
      \node [above] at (0,-0.4) {\tiny{$0$}};
      \end{tikzpicture}\\
\hline
\end{tabular}
\end{center}
\caption{Percolation-BEG coupling }
\label{ta2}
\end{table}

\begin{lemma}\label{lem43}
Let $Y(t)$ and $X(t)$ be the two Markov chains introduced above coupled according to the procedure of Table \ref{ta2}.
Consider the connected cluster $\gamma_{\rm BEG}$ of spins $+1$ containing the origin, if any, in the configuration $X(t)$, and
the connected cluster $\gamma_{\rm  Perc}$ of spins $+1$ containing the origin, if any, in the configuration $Y(t)$. Then
$$
\gamma_{\rm BEG}\subseteq\gamma_{\rm Perc}.
$$
\end{lemma}

\noindent{\bf Proof.}  The proof trivially follows from the fact that in the sampler of BEG model the probability to choose $\sigma_x=+1$ is, according Table \ref{ta2}, never bigger that $1/2$. \qed\\

\vglue.2truecm

\noindent{\bf Proof of Theorem \ref{teo41}.}
We may compute empirically by the strong law of large numbers, via the Markov chains $Y(t)$ and $X(t)$  coupled  as above,
the probabilities $\PP_\L^{\rm Perc}(o \rightsquigarrow\partial\L)$ and
$\PP_\L^+(o \rightsquigarrow\partial_e\L)$ where
 $\PP_\L^{\rm Perc}(o \rightsquigarrow\partial\L)$ is the probability that the origin is connected to the boundary of $\L$
by a connected set of open sites  in the two dimensional site percolation in $\L$ with parameter $p=1/2$ and
 $\PP_\L^+(o \rightsquigarrow\partial_e\L)$ is the probability that that the origin is connected to  $\partial_e\L$ by a connected set of $+$ in the zero-temperature two dimensional BEG model at FAD in $\L$ with $+$ boundary conditions.
 Lemma \ref{lem43} then immediately implies that
$\PP_\L^+(o\rightsquigarrow\partial_e\L)\le \PP^{\rm perc}_\L(o\rightsquigarrow\partial\L)$.

Now, it is well known (see e.g. Grimmett \cite{Gr}) that for two dimensional Bernoulli site percolation  the   value $p=1/2$ is subcritical hence
 $$
 \lim_{\L\uparrow \infty} \PP_\L^{\rm Perc}(o \rightsquigarrow\partial\L)=0
 $$
and thus
$$
\<\s_o\>^+=\lim_{\L\uparrow\infty} \PP_\L^+(o \rightsquigarrow\partial_e\L)\le
\lim_{\L\uparrow \infty}\PP_\L^{\rm Perc}(o \rightsquigarrow\partial\L)=0.
$$
Finally observe that  for any $\L$ we have by symmetry
$
\<\s_o\>_\L^-=-\<\s_o\>_\L^+
$
and thus Theorem \ref{teo41} follows.\qed

\subsection{Uniqueness of the Gibbs state of the zero-temperature BEG model at  FAD }\label{5}\mbox{}\\

In this section we prove the existence and  the independence on boundary conditions of the thermodynamic limit for the $n$-point
correlation functions, proving in this way the uniqueness of the Gibbs state.
Usually this kind of results are obtained by FKG inequalities (see e.g. \cite{sv17}), and it is possible to prove that FKG inequalities hold for the BEG model at FAD even at zero temperature \cite{LM,FKG}. We think however it is worth to provide here a proof of the uniqueness of the Gibbs state based solely on dynamical methods.
Notice also that e.g. at  the AD interface (i.e. $X=0$ and  $\YY<1$) the FKG inequalities are no longer satisfied so in principle the methods used here could be useful in situations where FKG inequalities cannot be used.

We will denote with $o_\L(1)$ any positive vanishing quantity in the thermodynamic limit, i.e., in the limit $L\to\infty$.
We will denote with $[n]=\{1,2,...,n\}$.

\noindent
Given $X$ and $Y$subsets of $\L\cup\partial_e\L$, we will denote with $X\rightsquigarrow Y$ the event (in the probability space $(F^\x_\L, \PP_\L^\x )$
in which at least one vertex in $X$ is connected with a path of vertices of non zero spins having the same sign to at least one vertex in $Y$. The complementary event will be denoted by $X\not\rightsquigarrow Y$.
Moreover, given  $X\subset \L$  and given   a fixed spin configuration   $\s_X$ in $X$ we denote
with the symbol  $\ev_{\s_X}$ the event  that in $X$ the spins are fixed at the configuration $\s_X$.
Given $X$ and $Y$subsets of $\L\cup\partial_e\L$, we will denote with $X\rightsquigarrow Y$ the event
(in the probability space $(\Si_\L, \PP_\L^\x )$)
in which at least one vertex in $X$ is connected with a path of vertices of non zero spins having the same sign to at least one vertex in $Y$. The complementary event will be denoted by $X\not\rightsquigarrow Y$.

Then  Theorem \ref{teo41} implies the following corollary on conditional probabilities in the BEG model.

\begin{corollary}\label{cor51}
Let $X,Y\subset \L$ be fixed subsets. Let $\x$ be any boundary condition on $\partial_e\L$ and let $\s_X$ be any fixed configurations of spins
in the sites of $X$. Then
\be\label{pp2}
\PP_\L^\xi(Y\rightsquigarrow\partial_e\L|\ev_{\s_X})=o_\L(1).
\ee
\end{corollary}

\noindent{\bf Proof}. We can proceed as in the proof of Theorem  \ref{teo41} using the coupling between site percolation
and BEG model described in Section \ref{4}. The only difference is that now the two coupled samplers
choose a site $x\in \L$ uniformly at random in the set $\L\setminus X$ and concerning the BEG model sampler
one must take into account that in the set $\L\setminus X$ the boundary conditions $\x$ are imposed on $\partial_e\L$ and the boundary
condition $\s_X$ are imposed at sites in $X$.
\qed

We will now state a result on $n$-point  correlations, and then we will prove the uniqueness of the Gibbs measure. In order to do that,
let us introduce some notations. In general, given  a collection  $\ev_1, \ev_2, \dots, \ev_n$ of   subsets of $F^\x_\L$ (i.e. events) we denote by $N^\x_\L|_{\ev_1,\dots,\ev_n}$ the number of ground states in $F^\x_\L$  such that the events $\ev_1, \ev_2, \dots, \ev_n$ occur.

\begin{theorem}\label{teo51} Given $x_1,x_2, \dots ,x_n\in \Z^2$ not necessarily distinct,
for any pair $\xi, \xi'$ of boundary conditions  we have
$$
\lim_{\L\uparrow \infty } [\langle\sigma_{x_1}...\s_{x_n}\rangle^{\xi}_\Lambda-
 \langle\sigma_{x_1}...\sigma_{x_n}\rangle^{\xi'}_\Lambda]=0.
 $$
\end{theorem}
\\{\bf Proof.}
\\Given  $I^-\subset [n]$, let   $I_+=[n]\backslash I_-$.
We denote shortly by $\ev_{I_+}$   the event $(\s_{x_i}=+1\ \forall i\in{I_+})$ and by $\ev_{I_-}$ the event $(\s_{x_i}=-1\ \forall i\in{I_-})$. Note that the event  $\ev_{I_+}$ is increasing while  $\ev_{I_-}$ is decreasing.  With these notations we can write
\begin{equation}\label{bou1}
\langle\sigma_{x_1}...\s_{x_n}\rangle^{\xi}_\Lambda=
\frac{1}{N_\L^\xi}\sum_{I_-\subset [n]}(-1)^{|I_-|}N^\x_\L|_{\ev_{I_+},\ev_{I_-}}
= \sum_{I_-\subset [n]}(-1)^{|I_-|}
\PP_\L^\x(\ev_{I_+}\cap\ev_{I_-}).
\end{equation}
Therefore  the theorem follows if we prove that for all choices of $I_-$ and for all pairs of boundary conditions $\xi,\xi'$ we have
\be\label{pint}
\lim_{\L\uparrow \infty}\left[\PP_\L^\x(\ev_{I_+}\cap\ev_{I_-})-\PP_\L^{\x'}(\ev_{I_+}\cap\ev_{I_-})\right]=0.
\ee

Observe that
$$
\PP_\L^\x(\ev_{I_+}\cap\ev_{I_-})=\PP_\L^\x(\ev_{I_+}|\ev_{I_-})\PP_\L^\x(\ev_{I_-}).
$$
So (\ref{pint}) follows  if we show that

$$
\lim_{\L\uparrow \infty } \left[\PP_\L^\x(\ev_{I_+}|\ev_{I_-}) - \PP_\L^{\x'}(\ev_{I_+}|\ev_{I_-})\right]=0\qquad \text{ and } \quad
\lim_{\L\uparrow \infty } \left[\PP_\L^\x(\ev_{I_-}) - \PP_\L^{\x'}(\ev_{I_-})\right]=0.
$$

Let us prove the first limit.
 The proof that $\lim_{\L\uparrow \infty } \left[\PP_\L^\x(\ev_{I_-}) - \PP_\L^{\x'}(\ev_{I_-})\right]=0$ is similar (and easier).

Since the event $\ev_{I_+}$ is increasing, for any boundary condition $\xi$ we have

$$
\PP^{-}_\L(\ev_{I_+}|\ev_{I_-})\le \PP^{\xi}_\L(\ev_{I_+}|\ev_{I_-})\le \PP^{+}_\L(\ev_{I_+}|\ev_{I_-}).
$$

\noindent
Indeed, setting  $X_{I_\pm}=\{x_i|i\in I_\pm\}$ and evaluating empirically the three probabilities as described in the proof of Corollary \ref{cor51} (i.e. the  coupled samplers are defined in $\L\setminus X_{I_-}$ with spins fixed at the value $-1$ in sites $X_{I_-}$).
Call now $S^\xi$ the set of $+$ obtained in the sampling with boundary condition $\xi$.
We have by order preserving coupling that $S^-\subset S^\xi\subset S^+$.

Now we can write
$$
\PP^{+}_\L(\ev_{I_+}|\ev_{I_-})=\PP^{+}_\L(\ev_{I_+}\cap(X_{I_+}\rightsquigarrow\partial_e\L)|\ev_{I_-})+
\PP^{+}_\L(\ev_{I_+}\cap(X_{I_+}{\not\rightsquigarrow}\partial_e\L)|\ev_{I_-}).
$$
By Corollary \ref{cor51} we get
$$
\PP^{+}_\L(\ev_{I_+}\cap(X_{I_+}\rightsquigarrow\partial_e\L)|\ev_{I_-})
\le \PP^{+}_\L(X_{I_+}\rightsquigarrow\partial_e\L|\ev_{I_-})=o_\L(1).
$$
Moreover

$$
\PP^{-}_\L(\ev_{I_+}|\ev_{I_-})\ge
\PP^{-}_\L(\ev_{I_+}\cap( X_{I_-}{\not\rightsquigarrow}\partial_e\L)|\ev_{I_-}).
$$
Hence we can write
$$
\PP^{-}_\L(\ev_{I_+}\cap( X_{I_-}{\not\rightsquigarrow}\partial_e\L)|\ev_{I_-})
\le P^{\xi}_\L(\ev_{I_+}|\ev_{I_-})\le
\PP^{+}_\L(\ev_{I_+}\cap(X_{I_+}{\not\rightsquigarrow}\partial_e\L)|\ev_{I_-}) +o_\L(1).
$$
The required independence of $\ P^{\xi}_\L(\ev_{I_+}|\ev_{I_-})$  from $\xi$ now follows from the following observation that
\be\label{diffe0}
\PP^{+}_\L(\ev_{I_+}\cap(X_{I_+}{\not\rightsquigarrow}\partial_e\L)|\ev_{I_-})-
\PP^{-}_\L(\ev_{I_+}\cap( X_{I_-}{\not\rightsquigarrow}\partial_e\L)|\ev_{I_-})
=o_\L(1).
\ee

Indeed, by definition
$$
\PP^{+}_\L(\ev_{I_+}\cap(X_{I_+}{\not\rightsquigarrow}\partial_e\L)|\ev_{I_-})=
{N^+_\L|_{\ev_{I_+},\ev_{I_-},X_{I_+}\not\rightsquigarrow\partial_e\L}\over N^+_\L|_{\ev_{I_-}}}
\dot{}$$
and
$$
\PP^{-}_\L(\ev_{I_+}\cap( X_{I_-}{\not\rightsquigarrow}\partial_e\L)|\ev_{I_-})=
{N^-_\L|_{\ev_{I_+},\ev_{I_-},X_{I_-}\not\rightsquigarrow\partial_e\L}\over N^+_\L|_{\ev_{I_-}}}{N^+_\L|_{\ev_{I_-}}\over N^-_\L|_{\ev_{I_-}}}.
$$
By spin flip symmetry we have
$N^-_\L|_{\ev_{I_+},\ev_{I_-},X_{I_-}\not\rightsquigarrow\partial_e\L}=
N^+_\L|_{\ev_{I_+},\ev_{I_-},X_{I_+}\not\rightsquigarrow\partial_e\L}$
so that
\be\label{diffe}
\PP^{+}_\L(\ev_{I_+}\cap(X_{I_+}{\not\rightsquigarrow}\partial_e\L)|\ev_{I_-})-
\PP^{-}_\L(\ev_{I_+}\cap( X_{I_-}{\not\rightsquigarrow}\partial_e\L)|\ev_{I_-})
=\PP^{+}_\L(\ev_{I_+}\cap(X_{I_+}{\not\rightsquigarrow}\partial_e\L)|\ev_{I_-})\Big[1-{N^+_\L|_{\ev_{I_-}}\over N^-_\L|_{\ev_{I_-}}}\Big].
\ee
Moreover, again by spin flip symmetry  we have that  $N^-_\L|_{\ev_{I_-},X_{I_-}\not\rightsquigarrow\partial_e\L}= N^+_\L|_{\ev_{I_-}}$ so that
$$
{N^-_\L|_{\ev_{I_-}}\over  N^+_\L|_{\ev_{I_-}}}
={N^-_\L|_{\ev_{I_-},X_{I_-}\not\rightsquigarrow\partial_e\L}+ N^-_\L|_{\ev_{I_-},X_{I_-}\rightsquigarrow\partial_e\L}\over N^+_\L|_{\ev_{I_-}}}
= 1+ {N^-_\L|_{\ev_{I_-},X_{I_-}\rightsquigarrow\partial_e\L}\over N^-_\L|_{\ev_{I_-}}}{N^-_\L|_{\ev_{I_-}}\over N^+_\L|_{\ev_{I_-}}}=1+o_\L(1){N^-_\L|_{\ev_{I_-}}\over N^+_\L|_{\ev_{I_-}}}.
$$
where in the last equality we have used that
${\N^-_\L|_{\ev_{I_-},X_{I_-}\rightsquigarrow\partial_e\L}\over N^-_\L|_{\ev_{I_-}}}=\PP^{-}_\L(X_{I_-}\rightsquigarrow\partial_e\L|\ev_{I_-})=o_\L(1)$  again by Corollary \ref{cor51}.
Therefore we get
\be\label{NN}
{N^+_\L|_{\ev_{I_-}}\over  N^-_\L|_{\ev_{I_-}}}=1-o_\L(1) .
\ee
Inserting (\ref{NN}) into (\ref{diffe}) we get (\ref{diffe0}).\qed

%
%
%

\vglue.2truecm
\noindent
We can now prove the following result.
\begin{theorem}\label{teo52}
For any choice of $x_1,...,x_n\in \L$ not necessarily distinct and for any boundary condition $\xi$ the limit
\be\label{slimit}
\lim_{\L\uparrow \infty } \langle\sigma_{x_1}...\s_{x_n}\rangle^{\xi}_\Lambda
\ee
 exists and it is independent of $\xi$. In other words, there is a unique infinite volume Gibbs measure  for the two-dimensional, zero-temperature BEG model at the FAD.
\end{theorem}

\noindent{\bf Proof.}   To prove that the limit (\ref{slimit}) is independent of $\x$ it is enough to prove,  by (\ref{pint}), that for all choices of $I_-$ and e.g. for boundary conditions $0$ the limit
\be\label{plimit}
\lim_{\L\uparrow \infty } \PP^{0}_\L(\ev_{I_+}\cap\ev_{I_-})
\ee
exists. Note  that using $0$ (which is equivalent to the free) boundary conditions  the event $X_{I_{\pm}}\rightsquigarrow\partial_e\L$ is empty.

Consider  two sets $\L$ and $\L'$ such that $\L'\subset\L$ and $X_{I_{\pm}}\subset\L'$.
Given a configuration $\s\in F^0_\L$, let $\s_{\partial_e\L'}$ be its restriction to $\partial_e\L'$ and let $U=\{\s_{\partial_e\L'}\}_{\s\in F^+_\L}$ the set of all feasible configurations $\x$ on $\partial_e\L'$.
Let moreover denote by $N^{0,\xi}_{\L\backslash\bar{\L}'}$ the number of ground states  in the ring $\L\setminus(\L'\cup\partial_e\L')$ compatible with boundary conditions $0$ on $\partial_e\L$ and $\xi$ on $\partial_e\L'$.
Then we can write
$$
\PP^{0}_\L(\ev_{I_+}\cap\ev_{I_-})=
{N^{0}_{\L}|_{\ev_{I_+},\ev_{I_-}}\over N^{0,\xi}_{\L}}=
\frac{\sum_{\xi\in U} N^{0,\xi}_{\L\setminus\bar{\L}'}
N^{\xi}_{\L'}|_{\ev_{I_+},\ev_{I_-}}}
{\sum_{\xi\in U} N^{0,\xi}_{\L\backslash\bar{\L}'}N^{\xi}_{\L'}}
=
\frac{\sum_{\xi\in U} N^{0,\xi}_{\L\backslash\bar{\L}' }N^{\xi}_{\L'}
\PP^\x_{\L'}(\ev_{I_+}\cap\ev_{I_-})}
{\sum_{\xi\in U} N^{0,\xi}_{\L\backslash\bar{\L}'}N^{\xi}_{\L'}}.
$$
By (\ref{pint}) we have that
$$
P^{\xi}_{\L'}(\ev_{I_+}\cap\ev_{I_-})=
P^{0}_{\L'}(\ev_{I_+}\cap \ev_{I_-})+o_{\L'}(1)
$$
and hence we have obtained
$$
\PP^{0}_\L(\ev_{I_+}\cap\ev_{I_-})=
\frac{\sum_{\xi\in U} N^{+,\xi}_{\L\backslash\bar{\L}' }N^{\xi}_{\L'}
[\PP^{0}_{\L'}(\ev_{I_+}\cap\ev_{I_-})+o_{\L'}(1)]}
{\sum_{\xi\in U} N^{+,\xi}_{\L\backslash\bar{\L}'}N^{\xi}_{\L'}}=
\PP^{0}_{\L'}(\ev_{I_+}\cap\ev_{I_-})+o_{\L'}(1).
$$
Therefore as $\L\uparrow\infty$ along a chosen sequence of square boxes $\{\L_n\}_{n\in \N}$, the sequence $\PP^{0}_{\L_n}(\ev_{I_+}\cap\ev_{I_-})$  is Cauchy and therefore the limit (\ref{plimit}) exists. Moreover this limit is
the same for all boundary conditions $\x$ because of (\ref{pint}).

\qed

\subsection{Exponential decay of the two point correlation for the zero-temperature BEG model at  FAD }\label{6}\mbox{}\\

We have shown in the previous section that $\<\s_x\s_y\>=\lim_{\L\uparrow\infty}\<\s_x\s_y\>_\L^\x$ exists and its is independent on the boundary condition $\x$. The main result of this section is stated in the following theorem.

\begin{theorem}\label{teo4}
There exist positive constants $C$ and $k$ such that
$$
\<\sigma_x\sigma_y\>\le C e^{-k|x-y|}.
$$
\end{theorem}

\noindent{\bf Proof}. We can write
$$
\< \s_{x}\s_y\>_\L^0=
{1\over   N^0_\L}\left[N^0_\L|_{\s_x\s_y=1}-N^0_\L|_{\s_x\s_y=-1}\right].
$$
Recalling that the event $\{x,y\}\rightsquigarrow\partial_e\L$ is empty under $0$ boundary conditions, let
$x\conplus y$ (resp. $x\conminus y$) be the event that $x$ and $y$ are in the same connected cluster of + spins (resp. - spins)
and let  $x\not\rightsquigarrow y$ be the event that $x$ and $y$ are in different  connected clusters, each of them with spins of the same sign.
With this definitions we have that
$$
N^0_\L|_{\s_x\s_y=1}= N^0_\L|_{x\conplus y } +  N^0_\L|_{x\conminus y }
+ N^0_\L|_{\s_x=\s_y=1, x\not\rightsquigarrow y }+ N^0_\L|_{\s_x=\s_y=-1, x\not\rightsquigarrow y }
$$
and
$$
N^0_\L|_{\s_x\s_y=-1}= N^0_\L|_{\s_x=-\s_y=-1, x\not\rightsquigarrow y }+
N^0_\L|_{\s_x=-\s_y=1, x\not\rightsquigarrow y }.
$$
Now observe that by spin flip symmetry
$$
N^0_\L|_{x\conplus y }= N^0_\L|_{x\conminus y }
$$
$$
N^0_\L|_{\s_x=\s_y=1, x\not\rightsquigarrow y }=
N^0_\L|_{\s_x=\s_y=-1, x\not\rightsquigarrow y }= N^0_\L|_{\s_x=-\s_y=-1, x\not\rightsquigarrow y }=
N^0_\L|_{\s_x=-\s_y=1, x\not\rightsquigarrow y }.
$$
Therefore
 we get
 $$
N^0_\L|_{\s_x\s_y=1}-N^0_\L|_{\s_x\s_y=-1}=2N^0_\L|_{x\conplus y }
 $$
 and hence
\be\label{pe13}
\< \s_{x}\s_y\>_\L^0=2\PP^0_\L(x\conplus y)
\ee

Taking thus  the limit $\Lambda\uparrow \infty$ in (\ref{pe13})(which, according to Theorem \ref{teo52}, exists and does not depend on boundary conditions)
we get
$$
\< \s_{x}\s_y\>=  \lim_{\L\uparrow\infty} 2\PP^0_\L(x\conplus y).
$$
Now, using the coupling BEG-Percolation introduced  in Lemma \ref{lem43} we get that
$$
\PP^+_\L(x\conplus  y)\le \PP^{\rm Perc}_\L(x\leftrightsquigarrow  y)
$$
where  $\PP^{\rm Perc}_\L(x\leftrightsquigarrow  y)$ is the probability that $x$ and $y$ are in the same connected open cluster in the two-dimensional Bernoulli site percolation  with parameter $p={1 \over 2}$.  It is well known (see e.g. \cite{Gr} and references therein) that, for some $k>0$ and $C'>0$,   $\PP^{\rm Perc}_\L(x\leftrightsquigarrow  y)\le C'e^{-k|x-y|}$. Hence we finally get
$$
\< \s_{x}\s_y\>\le  Ce^{-k|x-y|}
$$
with $C=2C'$.\qed

\section{Magnetization in the low temperature two-dimensional BEG model at the AD interface}\label{7}\mbox{}\\

In this section we will focus our attention on the AD interface of the BEG model in two dimensions confined in a box $\Lambda \subset \Z^2$ centered at the origin $o$ of $\Z^2$ with $+$ boundary conditions at finite inverse temperature $\beta$.
The Hamiltonian in this case is as follows.
\begin{equation}\label{hamad}
 H_{\Lambda,\YY}^+(\sigma)=-\sum_{x\sim y\atop \{x,y\}\cap \L\neq \0}\sigma_x\sigma_y(1+\YY\sigma_x\sigma_y)
 \end{equation}
where $\YY<-1$ and we agree that $\s_x=+1$ when $x\in \partial_e\L$.
The Gibbs mesure $\PP_{\L,\b}^+(\s)$ of a given configuration $\s\in \Si_\L$ is now  given by
\begin{equation}\label{gibbsad}
\PP_{\L,\b,\YY}^+(\s) = {e^{-\b H_{\Lambda,\YY}^+(\sigma)}\over Z^+_\L(\b,\rm \YY)},
\end{equation}
where
$$
{ Z^+_\L(\b,\rm {\YY})} = {\sum_{\s\in \Si_\L} }e^{-\b H_{\Lambda,\YY}^+(\sigma)}.
$$
As before, our goal is to evaluate the expected value $\<\s_o\>_{\L,\b,\YY}^+$ of the magnetization in the origin $o$ with respect to the Gibbs measure (\ref{gibbsad}), which is now given by
\begin{equation}\label{mago}
\<\s_o\>_{\L,\b,\YY}^+=\sum_{\s\in \Si_\L} \s_o\,\PP_{\L,\b,\YY}^+(\s),
\end{equation}
in the thermodynamic limit $\L\uparrow\infty$. The main result of this section is the following theorem.

\begin{theorem}\label{teoAD}
For the BEG model at the AD interface in $d=2$ and inverse temperature $\beta$, we have that
$$
\lim_{\L\uparrow\infty}\<\s_o\>_{\L,\b,\YY}^+=\lim_{\L\uparrow\infty}\<\s_o\>_{\L,\b,\YY}^-= 0,
$$
for any $\beta$ such that
\be\label{condiben}
e^{-\beta |\YY|}\cosh(\b) <  \frac{1-p_c}{2p_c},
\ee
where $p_c$ is the critical site percolation threshold in the square lattice.
\end{theorem}
\def\Le{\Lambda^{\rm e}}
\def\Lo{\Lambda^{\rm o}}
\def\so{\sigma^{\rm o}}
\def\se{\sigma^{\rm e}}
\def\tao{\tau^{\rm o}}
\def\te{\tau^{\rm e}}
\def\Siel{\Sigma_{\Le}}
\def\Siol{\Sigma_{\Lo}}

In order to prove the theorem above we need to prove a preliminary lemma analogous to Lemma \ref{lem42} of Section \ref{4}. \\Let thus $[o \rightarrow \partial\L]$ be the event formed by those configurations $\s\in \Si_\L$ for which there is a path $\ell$ of vertices connecting the origin $o$ to the boundary  $\partial_e\L$  in the original lattice $\L$  such that $|\s_x|=1$ for all $x\in \ell$. The complementary event will be denoted by $[o\not\rightarrow \partial\L]$.

\begin{lemma}\label{lem72}
The mean value of the spin in  the origin in the BEG model at the AD interface confined in a box $\L$ with $+$ boundary conditions
is given by
\begin{equation}\label{mag2}
\<\s_o\>_{\L,\b,\YY}^+
\leq
\PP_{\L,\b,\YY}^+([o \rightarrow \partial\L]),
\end{equation}
where  $\PP_\L^+$ is  Gibbs measure defined in (\ref{gibbsad}).
\end{lemma}

\noindent{\bf Proof.} Recalling the definition  of contours surrounding the origin given  in Sec. \ref{4},
similarly to what was remarked in the proof of Lemma \ref{lem42},  if $\s$ is a configuration in $\L$ belonging to the event  $[o\not\rightarrow \partial \L]$, then necessarily there is a unique
 contour $\g_\s$ surrounding the origin such that $\s_x=0$ for all $x\in \g_\s$ and such that the unique   contour contained in
the  subset $I_{\g_\s}^0(\s)\cup \g_\s$   is $\g_\s$.
Let us denote by $\Si^{\pm}_{I_\g}$
the set of all configurations  $\s$ in $I_\g$  such that the unique   contour in $I^0_{\g}(\s)$ is   $\g$ and such that $\s_o=\pm 1$.  Let
$\ev_o^\pm$ denotes the event that $\s_o=\pm 1$ and let us consider the quantity
\be\label{quanty}
\sum_{\s\in  [o\not\rightarrow\partial\L]} \s_o\,\PP_{\L,\b,\YY}^+(\s)=\sum_{\s\in  [o\not\rightarrow\partial\L]\atop\s_o=+1} \PP_{\L,\b,\YY}^+(\s)
-\sum_{\s\in  [o\not\rightarrow\partial\L]\atop\s_o=-1} \PP_{\L,\b,\YY}^+(\s).
\ee
Then, according to the notations above, we have that
\be\label{qua1}
\sum_{\s\in  [o\not\rightarrow\partial\L]\atop\s_o=\pm1} \PP_{\L,\b,\YY}^+(\s)=
{1\over Z^+_\L(\b,\YY)} \sum_{\s\in  [o\not\rightarrow\partial\L]\atop\s_o=\pm1}e^{-\b H_{\Lambda,\YY}^+(\sigma)}
=
{1\over Z^+_\L(\b,\YY)}
\sum_{\g\in \CC^o_\L}\sum_{\s\in\Si_\L\atop\s_o=\pm1, \g_\s=\g}e^{-\b H_{\Lambda,\YY}^+(\sigma)}.
\ee
Now observe that, for each minimal contour $\g\in \CC^o_\L$, we have
$$
\sum_{\s\in\Si_\L\atop\s_o=\pm1, \g_\s=\g}e^{-\b H_{\Lambda,\YY}^+(\sigma)}=
\sum_{\s\in\Si_{E^\L_\g}}e^{-\b H_{E^\L_\g,\YY}^+(\sigma)}\sum_{\s\in \Si^{\pm}_{I_\g}}e^{-\b H_{I_\g,\YY}^0(\sigma)},
$$
where given $\s\in\Si_{E^\L_\g}$,    $H_{E^\L_\g,\YY}^+(\sigma)$   is the  energy of $\s$ with boundary conditions $\s_x=+1$ when $x\in\partial_e\L$ and $\s_x=0$ when $x\in \g$, and, given $\s\in \Si^{\pm}_{I_\g}$, $H_{I_\g,\YY}^0(\sigma)$ is the energy
of the configurations $\s$ with boundary conditions $\s_x=0$ when $x\in \g$.
By the spin flip symmetry we have that  $H_{I_\g,\YY}^0(\sigma)= H_{I_\g,\YY}^0(-\sigma)$ and that $|\Si^{+}_{I_\g}|=|\Si^{-}_{I_\g}|$, which imples that
\be\label{qua2}
\sum_{\s\in \Si^{+}_{I_\g}}e^{-\b H_{\Lambda,\YY}^0(\sigma)}= \sum_{\s\in \Si^{-}_{I_\g}}e^{-\b H_{\Lambda,\YY}^0(\sigma)}
\ee
and thus
$$
\sum_{\s\in  [o\not\rightarrow\partial\L]\atop\s_o=-1} \PP_{\L,\b,\YY}^+(\s)=\sum_{\s\in  [o\not\rightarrow\partial\L]\atop\s_o=+1} \PP_{\L,\b,\YY}^+(\s).
$$
So we get that
\be\label{qu0}
\sum_{\s\in  [o\not\rightarrow\partial\L]} \s_o\,\PP_{\L,\b,\YY}^+(\s)=0,
\ee
In conclusion  we can bound from above the magnetization as follows:
\begin{equation}\label{eq:mag_ad}
\begin{aligned}
\<\s_o\>_\L^+ & =
\sum\limits_{\s\in [o \rightarrow \partial\L]}
\s_o \PP_{\L,\b,\YY}^+(\s)+ \sum\limits_{\s\in [o \not\rightarrow \partial\L]}
\s_o \PP_{\L,\b,\YY}^+(\s)\\
& = \sum\limits_{\s\in [o \rightarrow \partial\L]}
\s_o \PP_{\L,\b,\YY}^+(\s) \\
& =
\sum\limits_{\s\in[o \rightarrow \partial\L] \atop \s_o = +1}
\PP_{\L,\b,\YY}^+(\s) - \sum\limits_{\s\in[o \mapsto \partial\Lambda] \atop \s_o = -1} \PP_{\L,\b,\YY}^+(\s)\\
&\le
\sum\limits_{\s\in [o \rightarrow \partial\L]}
\PP_{\L,\b,\YY}^+(\s) = \PP_{\L,\b,\YY}^+([o \rightarrow \partial\L])\\
\end{aligned}
\end{equation}
and this concludes the proof of the lemma. $\Box$

\noindent {\bf Proof of Theorem \ref{teoAD}}. By Lemma \ref{lem72},  Theorem \ref{teoAD} is  proved  once we show that
\be\label{proba0}
\lim_{\L\uparrow\infty}\PP_{\L,\b,\YY}^+([o \rightarrow \partial\L])=0.
\ee

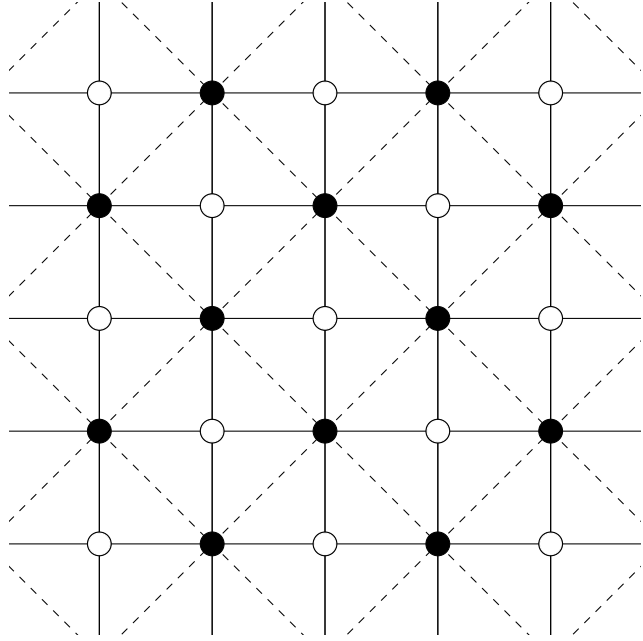
\begin{figure}
\centering
\begin{tikzpicture}[scale=1.5]
\clip (0.2,0.2) rectangle (5.8,5.8);
\foreach \x in {0,...,5}
  {\foreach \y in {0,...,5}
    {\draw[thin] (\x,\y) -- (\x,\y+1) -- ( \x+1,\y+1) -- (\x+1,\y);
     \filldraw[black] (\x,\y) circle (3pt);
   }
}
\draw[thin] (0,0) -- (6,0);
\foreach \x in {0,...,3}
  {\foreach \y in {0,...,3}
    {

    \draw[thin,dashed] (2*\x-1,2*\y) -- (2*\x+1,2*\y+2);
    \draw[thin,dashed] (2*\x-1,2*\y+2) -- (2*\x+1,2*\y);
     \filldraw[white] (2*\x+1,2*\y+1) circle (2.7pt);
    \filldraw[white] (2*\x,2*\y) circle (2.7pt);
   }
}
\end{tikzpicture}
\caption{AD LATTICE}
\label{fig:ad_lattice}
\end{figure}

To prove (\ref{proba0}), we  look for an upper   bound of $\PP_{\L,\b,\YY}^+([o \rightarrow \partial\L])$.  In order to do that we need to introduce some notations and definitions. Let $\mathbb{Z}^2_{\rm odd}$
(resp. $\mathbb{Z}^2_{\rm even}$) be the odd (resp. even) sublattices of $\mathbb{Z}^2$, i.e. $\mathbb{Z}^2_{\rm odd}$   (resp. $\mathbb{Z}^2_{\rm even}$)  is formed by those $(n_1,n_2)\in \mathbb{Z}^2$ such that $n_1+n_2$ is odd (resp. even). Let
$\Le=\L\cap\mathbb{Z}^2_{\rm even}$  and $\Lo=\L\cap\mathbb{Z}^2_{\rm odd}$. Note that the origin $o$ belongs to the sublattice $\Le$ and that
$\L= \Le\cup \Lo$ and $\Le\cap \Lo=\0$.
We  denote by  $\se$ ($\so$) a generic spin configuration  on  the even sublattice $\Le$ (on the odd sublattice $\Lo$), so that  (by a somehow abuse of notations) $\se\cup\so$ will denote a spin configuration on the original  lattice
$\L$ where $\sigma|_{\Le}=\se$ and $\sigma|_{\Lo}=\se$.
Given a configuration $\se$ in $\Le$
we recall that $\ev_{\se}$ denotes
the event formed by all configurations $\s\in \Si_\L$ such that $\s_z=\se_z$ for all $z\in \Le$.
We let $\Siel$ ($\Siol$) be the set of all spin configurations in  the even lattice $\Le$ (in  the odd lattice $\Lo$).
Observe that if $\{x,y\}\subset \Lo$ (or $\{x,y\}\subset \Le$) then necessarily
$|x-y|\ge 2$.
Let us denote by $\mathbb{G}_{\Lo}$ (resp. $\mathbb{G}_{\Le}$) the graph with vertex set $\Lo$ (resp. $\Le$) and edge set formed by the pairs $\{x,y\}\subset \Lo$ (resp. $\{x,y\}\subset \Le$) such that $|x-y|=2$. E.g., looking at Figure \ref{fig:ad_lattice}, the edges of the graph  $\mathbb{G}_\Lo$ are either  horizontal and vertical lines connecting two black (odd) sites passing through a white (even) site or   dashed diagonal lines indicated in Figure  \ref{fig:ad_lattice}. Therefore  $\mathbb{G}_{\Lo}$ (and similarly  $\mathbb{G}_{\Le}$) has as vertex  set  a subset of a square lattice $\mathbb{Z}^2$
in which each site is connected by an edge to 8 neighbors, namely  4 nearest neighbor (at Euclidean distance $\sqrt{2}$) and   4 next nearest neighbors (at Euclidean distance ${2}$).
Finally, given a spin configuration $\se$ on the
  even lattice $\Le$ we denote by $\mathbb{G}_{\L^{\rm o}}^{\se}$ the subgraph of $\mathbb{G}_{\Lo}$ with vertex set $\Lo$ and edge set formed by those pairs $\{x,y\}\in \Lo$  which are end points of a three-vertex path $\{x,z,y\}$ in $\L$ such that $|\s_z|=1$.

Given a configuration $\s=\se\cup\so \in\Si_\L$ and denoting for any  site $x\in \Lo$ by $\G_x$ the set if its neighbors (i.e. $\G_x=\{y\in \Le: |x-y|=1\}$),  we set
$$
H_x(\so_x|\se)= -\sum_{y\in \G_x}[\so_x\se_y-|\YY|(\so_x)^2(\se_y)^2].
$$

Notice that, by definition of conditional probability and due to the structure of the Hamiltonian (\ref{hamad}),  we have
$$
\PP_{\L,\b,\YY}^+(\so|\ev_{\se})=\frac{e^{-\beta H(\se\cup\so)}}{\sum_{\xi^{\rm o}}e^{-\beta H(\se \cup\xi^{(\rm o)})}}=
\prod_{x\in \Lo}
\left({e^{-\b H_x(\so_x|\se)}\over 1+\sum_{\x^{o}_x=\pm 1}e^{-\b H_x(\x^{\rm o}_x|\se)}}\right):=\prod_{x\in \Lo} \PP_{\L,\b,\YY}^+(\so_x|\ev_{\se}).
$$

We are now ready to bound $\PP_{\L,\b,\YY}^+([o \rightarrow \partial\L])$ from above. According to the notations previously introduced,  we  can write $\PP_{\L,\b,\YY}^+([o \rightarrow \partial\L])$ as follows.
\begin{equation}\label{eq:ad_1}
\begin{aligned}
\PP_{\L,\b,\YY}^+([o \rightarrow \partial\L])
&= \sum_{\s^{\rm e}\in \Siel} \PP_{\L,\b,\YY}^+([o \rightarrow \partial\L] \cap \ev_{\se})
\\
&= \sum_{\s^{\rm e}\in \Siel} \PP_{\L,\b,\YY}^+([o \rightarrow \partial\L] |\ev_{\se}) \PP_{\L,\b,\YY}^+(\ev_{\se}).
\end{aligned}
\end{equation}

\\Let  $[o{\xrightarrow{\mathbb{G}_{\Le}}\partial\L}]$
(resp. $[o{\xrightarrow{\mathbb{G}_{\Lo}}\partial\L}]$) be the event formed by those configurations $\se$ in $\Siel$ (resp. $\so$ in $\Siol$) for which the origin $o$  (resp. some neighbor of the origin $o$) is connected in $\mathbb{G}_{\Le}$ (in $\mathbb{G}_{\Lo}$)
to the boundary $\partial_e\L$ via  a path $\ell^{\rm e}$ ($\ell^{(\rm o}$) such that $|\se_x|=1$ for all $x\in \ell^{\rm e}$ ($|\so_x|=1$ for all $x\in \ell^{\rm o}$). Note that if $\se\cup\so\in[o\rightarrow\partial \L]$ then necessarily $\se\in[o{\xrightarrow{\mathbb{G}_{\Le}}\partial\L}]$ and  $\so\in[o{\xrightarrow{\mathbb{G}_{\Lo}}\partial\L}]$.

Let us also define $[o{\xrightarrow{\mathbb{G}_{\Lo}^{\se}}}\partial\L]$ the event  formed by all configurations $\so$ in $\Siol$ for which some neighbor of the origin $o$ is connected in the graph $\mathbb{G}_{\Lo}^{\se}$
to the boundary $\partial\L$ via  a path $\ell^{\rm o}$ such that $|\so_x|=1$ for all
$x\in \ell^{\rm o}$.
Note that, given $\se\in [o{\xrightarrow{\mathbb{G}_\Le}\partial\L}]$,  the graph  $\mathbb{G}_{\Lo}^{\se}$ is in general not connected
(as a subgraph of $\mathbb{G}_{\Lo})$. On the other hand,   for all  $\se\in [o{\xrightarrow{\mathbb{G}_\Le}\partial\L}]$, the graph
$\mathbb{G}_{\Lo}^{\se}$ has necessarily a  connected subgraph $\mathbb{C}_{\Lo}^{\se}$ with vertex set $V_{\Lo}^{\se}$ and edge set
$E_{\Lo}^{\se}$ such that $V_{\Lo}^{\se}$  contains a neighbor of the origin and a vertex  of $\partial_i\L\cap \Lo$. Moreover, for any
$\se\in  [o{\xrightarrow{\mathbb{G}_{\Le}}\partial\L}]$, each vertex $x\in V_{\Lo}^{\se}$ is such that $\sum_{y\sim x}|\se_y|\ge 1$.
By this observations,  we have

\begin{eqnarray*}\label{eq:ad_1.5}
\PP_{\L,\b,\YY}^+([o \rightarrow \partial\Lambda])
&\le&  \sum_{\s^{\rm e}\in \Siel} \PP_{\L,\b,\YY}^+([o \rightarrow \partial\Lambda] |\ev_{\se}) \PP_{\L,\b,\YY}^+(\ev_{\se})
\\
&=& \sum_{\s^{\rm e}\in [o{\xrightarrow{\mathbb{G}_{\Le}}\partial\L}]} \PP_{\L,\b,\YY}^+([o \rightarrow \partial\Lambda] |\ev_{\se}) \PP_{\L,\b,\YY}^+(\ev_{\se})
\\
&=&  \sum_{\s^{\rm e}\in [o{\xrightarrow{\mathbb{G}_{\Le}}\partial\L}]}
\PP_{\L,\b,\YY}^+([o{\xrightarrow{\mathbb{G}_{\Lo}^{\se}}}\partial\L]|\ev_{\se}) \PP_{\L,\b,\YY}^+(\ev_{\se})
\\
&\leq&  \sup_{\s^{\rm e}\in [o{\xrightarrow{\mathbb{G}_{\Le}}\partial\L}]} \PP_{\L,\b,\YY}^+([o{\xrightarrow{\mathbb{G}_{\Lo}^{\se}}}\partial\L]  |\ev_{\se})
\\
&=&  \sup_{\s^{\rm e}\in [o{\xrightarrow{\mathbb{G}_{\Le}}\partial\L}]}
\sum_{\so\in [o{\xrightarrow{\mathbb{G}_{\Lo}^{\se}}}\partial\L]}\PP_{\L,\b,\YY}^+(\so|\ev_{\se})
\\
&=&  \sup_{\s^{\rm e}\in [o{\xrightarrow{\mathbb{G}_{\Le}}\partial\L}]} \sum_{\so\in [o{\xrightarrow{\mathbb{G}_{\Lo}^{\se}}}\partial\L]}
\prod_{x\in \Lo}
\PP_{\L,\b,\YY}^+(\s_x^{\rm o}|\ev_{\se}).
\end{eqnarray*}

Given now $\so\in \Siol$, for any $x\in \Lo$ let us set $\omega_x=|\s^{\rm o}_x|$.   Let $\omega$
be a generic configuration in $\{0,1\}^{\Lo}\equiv\O_{\Lo}$, as in Section \ref{4}, we say that the site $x$ is open if $\omega_x=1$ and closed if
 $\omega_x=0$. Let $[o{\xrightarrow[\O_{\Lo}]{\mathbb{G}_{\Lo}^{\se}}}\partial\L]$ be the set of all configurations
$\omega\in \O_{\Lo}$ such that there exists a path in $\mathbb{G}_{\Lo}^{\se}$ of open sites connecting (a neighbor of) the origin to the boundary $\partial\L$.
Then we can rewrite

$$
 \sum_{\so\in [o{\xrightarrow{\mathbb{G}_{\Lo}^{\se}}}\partial\L]}
\prod_{x\in \Lo}
\PP_{\L,\b,\YY}^+(\s_x^{\rm o}|\ev_{\se})=
\sum_{\omega\in [o{\xrightarrow[\O^{\rm o}_\L]{\mathbb{G}_{\Lo}^{\se}}}\partial\L]}\prod_{x\in \Lo\atop \omega_x=1}p_x(\se) \prod_{x\in \Lo\atop \omega_x=0}(1-p_x(\se)),
$$
where
$$
p_x(\se)=  \PP_{\L,\b,\YY}^+(|\s_x^{\rm o}|=1 | \s^{\rm e}).
$$
Hence
\be
\begin{aligned}
\PP_{\L,\b,\YY}^+([o \mapsto \partial\Lambda]) &
\le \sup_{\s^{\rm e}\in [o{\xrightarrow{\mathbb{G}_{\Le}}\partial\L}]} \sum_{\so\in [o{\xrightarrow{\mathbb{G}_{\Lo}^{\se}}}\partial\L]}
\prod_{x\in \Lo}
\PP_{\L,\b,\YY}^+(\s_x^{\rm o}|\ev_{\se})\\
&=  \sup_{\s^{(\rm e)}\in [o{\xrightarrow{\mathbb{G}_\Le}\partial\L}]}
\sum_{\omega\in [o{\xrightarrow[\O^{\rm o}_\L]{\mathbb{G}_{\Lo}^{\se}}}\partial\L]}\prod_{x\in  \Lo\atop \omega_x=1}p_x(\se)
\prod_{x\in  \Lo\atop \omega_x=0}(1-p_x(\se))\\
&=  \sup_{\s^{(\rm e)}\in [o{\xrightarrow{\mathbb{G}_\Le}\partial\L}]}
\mathbb{P}^{\mathbb{G}_{\Lo}^{\se}}_{\{p_x(\se)\}_{x\in \Lo}}\Big([o{\xrightarrow[\O^{\rm o}_\L]{\mathbb{G}_{\Lo}^{\se}}}\partial\L]\Big),
\end{aligned}
\ee
where $\mathbb{P}^{\mathbb{G}_{\Lo}^{\se}}_{\{p_x(\se)\}_{x\in \Lo}}$ is the Bernoulli site percolation measure
in the graph $\mathbb{G}_{\Lo}^{\se}$,  where  each site $x\in \Lo$  is open with probability $p_x(\se)$ and closed with
probability $1-p_x(\se)$.

Recall now that, as said above,  the subgraph $\mathbb{G}_{\Lo}^{\se}$ of $\mathbb{G}_{\Lo}$ has a unique connected (in $\mathbb{G}_{\Lo}$) component $\mathbb{C}_{\Lo}^{\se}$
with vertex set $V_{\Lo}^{\se}$  containing the origin  $o$ and a site of the  boundary $\partial \L$.
This means that the event
$[o{\xrightarrow[\O^{\rm o}_\L]{\mathbb{G}_{\Lo}^{\se}}}\partial\L]$
 may only depend of the values of the $\omega_x$ with $x\in V_{\Lo}^{\se}$ and it is independent of $\omega_x$ for
 $x\in \Lo\setminus V_{\Lo}^{\se}$. Therefore for any $\se \in [o{\xrightarrow{\mathbb{G}_{\Lo}^{\se}}}\partial\L]$, we  have

\begin{eqnarray*}
\begin{aligned}
\mathbb{P}^{\mathbb{G}_{\Lo}^{\se}}_{\{p_x(\se)\}_{x\in \Lo}}([o{\xrightarrow[\O^{\rm o}_\L]{\mathbb{G}_{\Lo}^{\se}}}\partial\L])&=
\sum_{\omega\in [o{\xrightarrow[\O^{\rm o}_\L]{\mathbb{G}_{\Lo}^{\se}}}\partial\L]}\prod_{x\in  \Lo\atop \omega_x=1}p_x(\se)
\prod_{x\in  \Lo\atop \omega_x=0}(1-p_x(\se))\\
&=
\sum_{\omega\in \{0,1\}^{V_{\Lo}^{\se}}: \;[o{\xrightarrow[\O_{\Lo}]{\mathbb{G}_{\Lo}^{\se}}}\partial\L]}
\sum_{\omega\in \{0,1\}^{\Lo\setminus V_{\Lo}^{\se}}}
\prod_{x\in  \Lo\atop \omega_x=1}p_x(\se)
\prod_{x\in  \Lo\atop \omega_x=0}(1-p_x(\se))
\\
&=  \sum_{\omega\in \{0,1\}^{V_{\Lo}^{\se}}: \;[o{\xrightarrow[\O_{\Lo}]{\mathbb{G}_{\Lo}^{\se}}}\partial\L]}
\prod_{x\in  V_{\Lo}^{\se}\atop \omega_x=1}p_x(\se)
\prod_{x\in  V_{\Lo}^{\se}\atop \omega_x=0}(1-p_x(\se)) \times
\\
& ~~~~\times  \sum_{\omega\in \{0,1\}^{\Lo\setminus V_{\Lo}^{\se}}}
\prod_{x\in  \Lo\setminus V_{\Lo}^{\se}\atop \omega_x=1}p_x(\se)
\prod_{x\in  \Lo\setminus V_{\Lo}^{\se}\atop \omega_x=0}(1-p_x(\se))\\
&= \sum_{\omega\in \{0,1\}^{V_{\Lo}^{\se}}: \;[o{\xrightarrow[\O_{\Lo}]{\mathbb{G}_{\Lo}^{\se}}}\partial\L]}
\prod_{x\in  V_{\Lo}^{\se}\atop \omega_x=1}p_x(\se)
\prod_{x\in  V_{\Lo}^{\se}\atop \omega_x=0}(1-p_x(\se)),
\end{aligned}
\end{eqnarray*}
since
$$
\sum_{\omega\in \{0,1\}^{\Lo\setminus V_{\Lo}^{\se}}}
\prod_{x\in  \Lo\setminus V_{\Lo}^{\se}\atop \omega_x=1}p_x(\se)
\prod_{x\in  \Lo\setminus V_{\Lo}^{\se}\atop \omega_x=0}(1-p_x(\se))=1 .
$$
Therefore, we have that
$$
\mathbb{P}^{\mathbb{G}_{\Lo}^{\se}}_{\{p_x(\se)\}_{x\in \Lo}}([o{\xrightarrow[\O_{\Lo}]{\mathbb{G}_{\Lo}^{\se}}}\partial\L])=
\mathbb{P}^{\mathbb{C}_{\Lo}^{\se}}_{\{p_x(\se)\}_{x\in V_{\Lo}^{\se}}}([o{\xrightarrow[V_{\Lo}^{\se}]{\mathbb{C}_{\Lo}^{\se}}}\partial\L])
$$
Moreover, recalling that for any
$\se\in  [o{\xrightarrow{\mathbb{G}_\Le}\partial\L}]$ each vertex $x\in V_{\Lo}^{\se}$ is such that $\sum_{y\sim x}|\se_y|\ge 1$, and setting
\be\label{okkk}
p \doteq \sup_{\sum_{y\sim x}|\se_y|\ge 1}p_x(\se),
\ee
we can bound
$$
\mathbb{P}^{\mathbb{C}_{\Lo}^{\se}}_{\{p_x(\se)\}_{x\in V_{\Lo}^{\se}}}([o{\xrightarrow[V_{\Lo}^{\se}]{\mathbb{G}_{\Lo}^{\se}}}\partial\L])\le
\mathbb{P}^{\mathbb{C}_{\Lo}^{\se}}_{p}([o{\xrightarrow[V_{\Lo}^{\se}]{\mathbb{C}_{\Lo}^{\se}}}\partial\L],
$$
 where now $\mathbb{P}^{\mathbb{C}_{\Lo}^{\se}}_{p}$ is the homogeneous Bernoulli site percolation probability measure
in the graph $\mathbb{C}_{\Lo}^{\se}$ with parameter $p$  given by (\ref{okkk}).

So we get that
\begin{equation}\label{eros}
\begin{aligned}
\PP_{\L,\b,\YY}^+([o \mapsto \partial\Lambda])
& \le  \sup_{\s^{(\rm e)}\in [o{\xrightarrow{\mathbb{G}_\Le}\partial\L}]}
\mathbb{P}^{\mathbb{C}_{\Lo}^{\se}}_{p}([o{\xrightarrow[V_{\Lo}^{\se}]{\mathbb{C}_{\Lo}^{\se}}}\partial\L])\\
& \le \mathbb{P}^{\mathbb{G}_{\Lo}}_{p}([o{\xrightarrow{\mathbb{G}_{\Lo}}}\partial\L])
\end{aligned}
\end{equation}
where $\mathbb{P}^{\mathbb{G}_{\Lo}}_{p}$ is the Bernoulli site percolation probability in the graph $\mathbb{G}_{\Lo}$.
The last inequality in  (\ref{eros}) follows from the fact that for any $\se\in [o{\xrightarrow{\mathbb{G}_\Le}\partial\L}]$ the
connected component $\mathbb{C}_{\Lo}^{\se}$ is a subgrah $\mathbb{G}_{\Lo}$  and the probability that the origin is connected to the boundary in any subgraph  $C$ of   $\mathbb{G}_{\Lo}$ is always less than (or at most equal to)  the probability that the origin is connected to the boundary
in the whole graph  $\mathbb{G}_{\Lo}$. Therefore  in (\ref{eros})
$\mathbb{P}^{\mathbb{C}_{\Lo}^{\se}}_{p}([o{\xrightarrow[V_{\Lo}^{\se}]{\mathbb{C}_{\Lo}^{\se}}}\partial\L])$ attains
the  supremum  over the configurations ${\s^{\rm e}\in [o{\xrightarrow{\mathbb{G}_\Le}\partial\L}]}$ such that $|\se_x|=1$ for all $x\in \Le$, i.e. those $\se$ for which   $\mathbb{C}_{\Lo}^{\se}= \mathbb{G}_{\Lo}$.

Let now $\mathbb{G}_1$  be the graph whose vertices set is $\mathbb{Z}^2$ and whose edges set are $\{x,y\}\in \mathbb{Z}^2$ such that
$||x-y||_\infty=1$.
It is well known (see e.g. \cite{Gr} and references therein) that the site percolation threshold $p^*_c$ of
  the  8-neighbor square lattice $\mathbb{G}_1$ defined above  is $p^*_c=1-p_c$ where $p_c$ is the site percolation threshold in the usual square lattice $\mathbb{Z}^2$ (with 4 neighbors). It is also well known via numerical simulations that that $p_c=0.592...$ so that $1-p_c=0.407...$ (see again \cite{Gr} and also \cite{MG}).
Let
moreover $\mathbb{G}^{\rm o}$ be the graph whose vertices set is $\mathbb{Z}^2_{\rm odd}$ and set of edges  those  $\{x,y\}\in \mathbb{Z}^2_{\rm odd}$ such that  $||x-y||_\infty=2$. Then ${ \mathbb{G}}_1$ and  $\mathbb{G}^{\rm o}$ are isomorphic so that  $\mathbb{P}^{{ \mathbb{G}}^{\rm o}}_{p}([o \xrightarrow{} \infty])=\mathbb{P}^{{ \mathbb{G}}_1}_{p}([o \xrightarrow{} \infty])$.   Also,  the graph   $\mathbb{G}_{\Lo}$ previously introduced is the restriction
of $\mathbb{G}^{\rm o}$ to $\Lo$ and therefore $\mathbb{G}_{\Lo}\to \mathbb{G}^{\rm o}$ as $\L\to\infty$.

Therefore, by (\ref{eros}) and Lemma \ref{lem72},  we get
  $$
\lim_{\L\to \infty} \langle\s_o\rangle^+_\L=\lim_{\L\to \infty} \PP_{\L,\b,\YY}^+([o \mapsto \partial\Lambda])\le
\lim_{\L\to \infty}\mathbb{P}^{\mathbb{G}_{\Lo}}_{p}([o{\xrightarrow{\mathbb{G}_{\Lo}}}\partial\L])=\mathbb{P}^{\mathbb{G}^{\rm o}}_{p}([o \xrightarrow{} \infty])=\mathbb{P}^{{ \mathbb{G}}_1}_{p}([o \xrightarrow{} \infty]).
$$

By means of a simple calculation, we will show below that
$p$ defined in (\ref{okkk}) is given by
\be\label{ppii}
p= \frac{2e^{-\beta|\YY|}\cosh(\beta )}{1+2e^{- \beta |\YY|}\cosh(\beta )}.
\ee

In conclusion,  recalling that  $p<(1-p_c)$  is subcritical for the site percolation in $\mathbb{G}_1$ and recalling (\ref{ppii}), we get that $\lim_{\L\to \infty} \langle\s_o\rangle^+_\L=0,$ whenever
$$
\frac{2e^{-\beta |\YY|}\cosh(\beta )}{1+2e^{- \beta |\YY|}\cosh(\beta )}<1-p_c,
$$
which is equivalent go (\ref{condiben}) and, by spin flipping symmetry,
 $\<\s_o\>_{\L,\b,\YY}^-=-\<\s_o\>_{\L,\b,\YY}^+=0$,
and  this concludes the proof of Theorem \ref{teoAD}.\qed

\mbox{}\\\

Next we will prove (\ref{ppii}). Let $x \in \Lo$ and suppose  $k\geq 1$ is the number of nonzero spins in $\G_x$. Let their sum be denoted by $N_k$, then
$N_1\in \{-1,1\}$, $N_2\in \{-2,0,2\}$, $N_3\in \{-3,-1, 1, 3\}$ and $N_4\in \{-4,-2,2,4\}$.
Notice that $$P(|\s_x|=1|\s^{{\rm e}})=P(\s_x=1|\s^{{\rm e}})+P(\s_x=-1|\s^{{\rm e}})=\frac{2e^{-k \beta|\YY|}\cosh(\beta N_k)}{1+2e^{-k \beta |\YY|}\cosh(\beta N_k)},$$
which is even in $N_k$, so we may restrict ourselves to non-negative values of $N_k$. For each fixed $k$, $P(|\s_x|=1|\s^{{\rm e}})$ is non-decreasing in $N_k$, besides, $N_k\leq k$, then
$$ P(|\s_x|=1|\s^{{\rm e}})\leq \frac{2e^{-k\beta |\YY|}\cosh(\beta k)}{1+2e^{-k \beta |\YY|}\cosh(\beta k)}.$$
Since $|Y|>1$, then $\frac{2e^{-k\beta |\YY|}\cosh(\beta k)}{1+2e^{-k \beta |\YY|}\cosh(\beta k)}$ is non-incresing in $k$ and since $k\geq 1$, we are done.


To conclude this section, let us compare our  condition (\ref{condiben}), which can be rewritten as
\begin{eqnarray}\label{condnew}\frac{2p_c}{1-p_c}\, e^{-\beta |\YY|}\cosh(\beta) <1,\end{eqnarray} with  the condition for vanishing of the magnetization for the
two dimensional BEG model at the  AD line obtained in \cite{pcl20} via  expansion methods, namely,
\be\label{paulo}
64e^{-\beta|\YY|}\sinh(\beta)<1.
\ee
A simple computation shows that, as long as,
$$\beta \geq  1/2 \ln \left(\frac{32(1-p_c)+1}{32(1-p_c)-1}\right)\approx 0.0783,$$
then
\begin{eqnarray*}\frac{2p_c}{1-p_c} e^{-\beta |\YY|} \cosh \beta\leq 64 e^{ -\beta |\YY|}\sinh(\beta),\end{eqnarray*}
for all $|\YY|>1$, in particular, for such values of $\beta$, the condition (\ref{condnew}) is better than (\ref{paulo}).

%
%

\section{Conclusions and open problems}\label{8}\mbox{}\\

In this paper we focus our attention on the zero-temperature BEG model at the FAD and AD  interfaces. We first perform some simulations on the two-dimensional and three-dimensional cases via  a perfect sampling of  an ergodic and
symmetric Markov Chain whose stationary measure
coincides with the zero-temperature Gibbs measure (uniform on the ground states). Our numerical data indicates that
the magnetization of the model is zero in the two-dimensional case and its is different from zero in the three-dimensional case.
These data also indicate that the two-point correlation decays exponentially at large distance
in the two-dimensional case. Next, we obtain several rigorous results about the  zero-temperature BEG  model at FAD in two dimensions. Namely, we prove rigorously, using dynamical arguments, that the mean value of the spin at the origin is zero in the two-dimensional case and we use this result together with some additional dynamical arguments to conclude the existence and uniqueness of the zero-temperature Gibbs measure for the two-dimensional BEG model at FAD and to prove rigorously that the two-point correlation decays exponentially fast at large distances.
Finally we prove, via comparison with Bernoulli site percolation, the  absence of magnetization of the BEG model in the whole AD line at low temperatures, such a condition improves earlier results obtained in \cite{pcl20}, via expansions.

Several interesting open problems can be addressed in the next future.
Concerning  the two dimensional BEG model at FAD, one could try to prove rigorously (maybe using cluster expansion methods) that  for $\b$ large but finite the magnetization is still zero as indicated by the  Monte Carlo simulations presented in \cite{rb04}.
{ We also point out  that
the uniqueness in $d=2$ of the BEG Gibbs measure at FAD also for large but finite $\b$ shows that the two-dimensional behaviour is very different from   the phase diagram shown in
Figure 1 (f) of reference \cite{bh91} obtained via mean field methods. Indeed,  according to this diagram,
as soon as the temperature is turned on, the BEG model at FAD should be in the ferromagnetic phase}.

Again about the two dimensional BEG model at FAD, it could be interesting to obtain tighter rigorous bounds
on the exponential decay of the two-point correlation function at zero temperature.

Concerning the three-dimensional case it remains completely open  to prove rigorously (possibly using dynamical arguments?) that the Gibbs measure is not unique at zero temperature.

\section*{Acknowledgments.}
Several discussions with many colleagues have been very useful during this work. We thank Lorenzo Bertini, Paolo Buttà, Emilio Cirillo, Alessandro Giuliani, Lucio Russo  and Elisabetta Scoppola.
This work is dedicated to Paolo Dai Pra in the occasion of his 60-th birthday.
R.M. and B.S. acknowledges the MIUR Excellence Department Project awarded to the Department of Mathematics, University of Rome Tor Vergata, CUP E83C18000100006.
 A. P. has been partially supported by the Brazilian science foundations Conselho Nacional de Desenvolvimento
Cient\'\i fico e Tecnol\'ogico (CNPq), Coordena\c{c}\~ao de Aperfeiçoamento de Pessoal de N\'\i vel Superior (CAPES) and Funda\c{c}\~ao de Amparo a Pesquisa do Estado de Minas Gerais (FAPEMIG), and by the University of Tor Vergata.

\end{document}